\documentclass[11pt,dvips]{article}
\usepackage{amscd,verbatim}
\usepackage[all]{xy}
\usepackage{graphicx}

\usepackage{amsmath}
\usepackage[psamsfonts]{amssymb}
\usepackage{amsthm}
\usepackage{euscript}

\usepackage{latexsym}

\renewcommand{\(}{\begin{equation}}
\renewcommand{\)}{end{equation} \vspace{-.05in}\linebreak}

\newcounter{saveeqn}
\newcounter{savealpheqn}

\newcommand{\alpheqn}{\setcounter{saveeqn}{\value{equation}}%
 \stepcounter{saveeqn}\setcounter{equation}{0}%
 \renewcommand{\theequation}{\mbox{\arabic{section}.\arabic{saveeqn}
\alph{equation}}}
 \renewcommand{\)}{\end{equation}}}
\def\part#1{\frac{\partial}{\partial{#1}}}%
\def\group#1{\refstepcounter{equation}\setcounter{saveeqn}{\value{equation}}%
 \label{#1}\setcounter{equation}{0}%
\renewcommand{\theequation}{\mbox{\arabic{section}.\arabic{saveeqn}
\alph{equation}}}
 \renewcommand{\)}{\end{equation}}}
\newcommand{\reseteqn}{\setcounter{equation}{\value{saveeqn}}%
 \renewcommand{\theequation}{\arabic{section}.\arabic{equation}}%
 \renewcommand{\)}{\end{equation}}}

\newcommand{\aalpheqn}{\setcounter{saveeqn}{\value{equation}}%
 \stepcounter{saveeqn}\setcounter{equation}{0}%
 \renewcommand{\theequation}{\mbox{
       \Alph{subsection}.\arabic{saveeqn}\alph{equation}}}
  \renewcommand{\)}{\end{equation}}}
\newcommand{\areseteqn}{\setcounter{equation}{\value{saveeqn}}%
 \renewcommand{\theequation}{\Alph{subsection}.\arabic{equation}}%
 \renewcommand{\)}{\end{equation}}}

\renewcommand{\thefootnote}{\alph{footnote}}
\renewcommand{\(}{\begin{equation}}
\renewcommand{\)}{\end{equation}}
\newcommand{\ba}{\begin{eqnarray}}
\newcommand{\ea}{\end{eqnarray}}

\newcommand{\bp}{\mathop{\vtop{\ialign{##\crcr
  $\hfil\displaystyle{}\hfil$\crcr\noalign{\kern-13pt\nointerlineskip}
  \BIG{(}\hskip0pt\crcr\noalign{\kern3pt}}}}}
\newcommand{\cbp}{\mathop{\vtop{\ialign{##\crcr
  $\hfil\displaystyle{}\hfil$\crcr\noalign{\kern-13pt\nointerlineskip}
  \BIG{)}\hskip0pt\crcr\noalign{\kern3pt}}}}}
\newcommand{\pa}{\mathop{\vtop{\ialign{##\crcr
  $\hfil\displaystyle{\oplus}\hfil$\crcr\noalign{\kern+1pt\nointerlineskip}
  \hspace{.08in}$^{\alpha=0}$\hskip6pt\crcr\noalign{\kern3pt}}}}}
\renewcommand{\sp}{,\hspace{.3in}}
\newcommand{\p}{^\prime}

\newcommand{\R}{\ensuremath{\mathbb R}}
\newcommand{\rpn}{\ensuremath{{\mathbb R {\text{P}}^n}}}
\newcommand{\rpt}{\ensuremath{{\mathbb R {\text{P}}^2}}}
\newcommand{\cB}{\ensuremath{\mathcal B}}
\newcommand{\cE}{\ensuremath{\mathcal E}}
\newcommand{\cF}{\ensuremath{\mathcal F}}

\newcommand{\cQ}{\ensuremath{\mathcal Q}}

\newcommand{\cL}{\ensuremath{\mathcal L}}
\newcommand{\cP}{\ensuremath{\mathcal P}}

\newcommand{\Z}{\ensuremath{\mathbb Z}}

\def\dwn{\downarrow}

\def\L{\ensuremath{{\cal L}}}

\newcommand{\beq}{\begin{equation}}
\newcommand{\eeq}{\end{equation}}

\numberwithin{equation}{section}

\def\hsp#1{\hspace{#1in}}

\catcode`\@=11
\def\vereq#1#2{\lower3pt\vbox{\baselineskip1.5pt \lineskip1.5pt
\ialign{$\m@th#1\hfill##\hfil$\crcr#2\crcr\sim\crcr}}}
\catcode`\@=12

\makeatletter
\newcommand\figcaption{\def\@captype{figure}\caption}
\newcommand\tabcaption{\def\@captype{table}\caption}
\makeatother
\renewcommand{\(}{\begin{equation}}
\renewcommand{\)}{\end{equation}}

\newcommand{\CC}{{\mathbb C}}
\newcommand{\RR}{{\mathbb R}}
\newcommand{\ZZ}{{\mathbb Z}}

\theoremstyle{plain}
\newtheorem{theorem}{Theorem}[section]
\newtheorem{lemma}[theorem]{Lemma}

\theoremstyle{definition}

\newcommand{\two}{\text{I}\!\text{I}}
\newcommand{\twoa}{\text{I}\!\text{IA}}
\newcommand{\twob}{\text{I}\!\text{IB}}
\newcommand{\CP}{\CC \text{P}}
\newcommand{\RP}{\RR \text{P}}
\newcommand{\rp}{\RP}

\begin{document}

% =========================================================================
\begin{titlepage}
\begin{flushright}
ADP-03-122/M102

IFUP-TH/2003/19

hep-th/0306062
\end{flushright}

\vspace{2em}
\def\thefootnote{\fnsymbol{footnote}}

\begin{center}
{\Large\bf T-Duality: Topology Change from $H$-flux}
\end{center}
\vspace{1em}

\begin{center}
Peter Bouwknegt\footnote{E-Mail:
 pbouwkne@physics.adelaide.edu.au, pbouwkne@maths.adelaide.edu.au}$^{1,2}$, 
Jarah Evslin\footnote{E-Mail: jarah@df.unipi.it}$^3$, 
and Varghese Mathai\footnote{E-Mail: vmathai@maths.adelaide.edu.au}$^2$\ 
\end{center}

\begin{center}
\vspace{1em}
{\em $^1$Department of Physics\\
     School of Chemistry and Physics \\
     University of Adelaide\\
     Adelaide, SA 5005, Australia}\\
\hsp{.3}\\
{\em $^2$Department of Pure Mathematics\\
     University of Adelaide\\
     Adelaide, SA 5005, Australia}\\
\hsp{.3}\\
{\em $^3$INFN Sezione di Pisa\\
     Via Buonarroti, 2, Ed.~C,\\
     56127 Pisa, Italy}\\

\end{center}

\vspace{0em}
\begin{abstract}
\noindent
T-duality acts on circle bundles by
exchanging the first Chern class with the fiberwise integral of the
$H$-flux, as we motivate using $E_8$ and also using $S$-duality.
We present known and new examples including NS5-branes, nilmanifolds,
Lens spaces, both circle bundles over $\rpn$, and the $AdS^5\times
S^5$ to $AdS^5\times \CP^2\times S^1$ with background $H$-flux of Duff,
L\"u and Pope.  When T-duality leads to M-theory on a non-$spin$ manifold the
gravitino partition function continues to exist due to the background
flux, however the known quantization condition for $G_4$ receives a correction.  In a
more general context, we use {\em correspondence spaces} to implement
isomorphisms on the twisted K-theories and twisted cohomology theories
and to study the corresponding Grothendieck-Riemann-Roch theorem.
Interestingly, in the case of decomposable twists, both twisted
theories admit fusion products and so are naturally rings.
\end{abstract}

\vfill

\end{titlepage}
\setcounter{footnote}{0} 
\renewcommand{\thefootnote}{\arabic{footnote}}

\pagebreak
% =========================================================================
\renewcommand{\thepage}{\arabic{page}}

\section{Introduction}

T-duality is a generalization of the $R\to 1/R$ invariance of string
theory compactified on a circle of radius $R$.  The local
transformation rules of the low energy effective fields under
T-duality, known as the Buscher rules \cite{Bus} (see also, e.g.,
\cite{RV,AAL,BHO}), have been known for some time, but global issues,
in particular in the presence of NS 3-form $H$-flux, have remained
obscure.  It is known, however, through many examples in the
literature \cite{AABL,DLP,GLMW,KSTT}, that the general case involves a
change in the topology of the manifold.  However no systematic method
has been developed for determining the topology change.  In this paper
we will propose a formula for the topology change under T-duality, and
we will show that it yields the desired isomorphism both in the
context of twisted cohomology as well as twisted K-theory.  We
conjecture that the duality holds, however, in the full string theory
as well.

To simplify the discussion we will restrict ourselves in this paper to
T-duality in one direction only, $i.e.$ T-dualizing on a circle $S^1$.
A more general case with a $d$-dimensional torus can be obtained by
successive dualizations so long as the integral of $H$ over each
2-subtorus vanishes.  If this integral does not vanish, then after
T-dualizing about one circle the other circle no longer exists.  We
will relate the obstruction to T-duality to a particular type of failure
of the 2-torus to lift to F-theory.\footnote{However there are torii that
do not lift to F-theory on which we may T-dualize, for example, a 2-torus
that supports $G_3$ flux.  S-dualizing, the obstruction to T-duality on
a torus with $H$-flux is the controversial obstruction to S-duality 
in the presence of $G_1$ flux.}  In
integral cohomology the story is the same, as the integral of $H$
inhabits $H^1(M,\Z)$ which cannot have a torsion piece because of the
Universal Coefficient Theorem.

First, consider the case where spacetime $E$ is a product manifold 
$M\times S^1$ and the NS 3-form $H$ is trivial in $H^3(E,\ZZ)$, $i.e.$
we can write $H=dB$ globally.  Similarly, for the T-dual
we have $\hat H=d\hat B$.  In this case, upon 
T-dualizing on $S^1$, the Buscher rules on the 
RR fields can be conveniently encoded in the formula \cite{Hor}
\begin{equation} \label{eqAa}
  \hat G = \int_{S^1}\ e^{\cF-B+\hat B}\ G \,,
\end{equation}
where $G$ is the total (gauge invariant) RR fieldstrength,
$G=\sum_p G_{p+2}$ ($p=0,2,4,\ldots,8$ for type $\twoa$ and
$p=-1,1,\ldots,7$ for type $\twob$), and $\cF=d\theta\wedge
d\hat\theta$ is the curvature 
of the Poincar\'e linebundle $\cP$
on $S^1 \times \hat S^1$, so that $e^\cF = ch(\cP)$ is the Chern 
character of $\cP$.  The right hand side of \eqref{eqAa}
is interpreted as a (closed) form on $M\times S^1 \times \hat S^1$,
and integrated along $S^1$ to yield a form on the T-dual space
$\hat E = M \times \hat S^1$.\footnote{Strictly speaking,
the various forms entering \eqref{eqAa} are the pull-backs 
of forms to the correspondence space $M\times S^1 \times \hat S^1$.}

The RR field $G$ is $d_H$-closed, where $d_H=d-H\wedge$ is the 
$H$-twisted differential, and it follows that its T-dual $\hat G$
is $d_{\hat H}$-closed.  This is just the supergravity Bianchi identity.
Gauge invariance is
implemented through $\delta C = e^B d\alpha$, where the gauge
potential $C$ is related to $G$ by
$G = e^B d(e^{-B} C) = d_H C$.  Thus, we can interpret \eqref{eqAa}
as an isomorphism
\begin{equation} \label{eqAb} \begin{CD}
  T_* : H^\bullet (M\times S^1,H) @>\cong>> 
  H^{\bullet+1} (M\times \hat S^1,\hat H). \end{CD}
\end{equation}
Of course, since in this case $H=dB$ globally, the twisted cohomology
$H^\bullet (E,H)$ is canonically isomorphic to the usual cohomology
$H^\bullet(E)$, by noting that $d(e^{-B}G) = e^{-B}d_H G$.  

The discussion above can be lifted to K-theory, \cite{Hor} (see also
\cite{Guk,Sha,OS,MoSa}), and thus to the classification of D-branes on
$M\times S^1$ and $M\times \hat S^1$, by using the correspondence
\begin{equation} \label{eqAc}
\xymatrix @=4pc { &  M\times S^1 \times \hat S^1 \ar[dl]^{p} 
\ar[dr]_{\hat p} & \\
M\times S^1   &   &    M\times \hat S^1 }
\end{equation}
This gives rise to an isomorphism of K-theories
\begin{equation} \label{eqAd} \begin{CD}
  T_! : K^\bullet(M\times S^1) @>\cong>> K^{\bullet+1}(M\times 
  \hat S^1) \end{CD}
\end{equation}\label{eqAe}
by
\begin{equation}
  T_! = \hat p_!\ ( p^!(\ \cdot\ )\, \otimes \, \cP ) \,.
\end{equation}

It is well-known that the application 
of T-duality is not restricted to product manifolds 
$M\times S^1$, but can also be applied locally in the case
of $S^1$-fibrations over $M$ \cite{SYZ}, and moreover, can be 
generalized to situations with nontrivial NS 3-form flux $H$.
While in this more general case, strictly speaking, \eqref{eqAa} 
does not make sense since neither the Poincar\'e bundle, nor $B$,
are defined globally, it does appear that in some sense the 
equation still makes sense locally as it does give 
rise to the correct Buscher rules even in this more general setting.

In this paper we investigate the more general case where $E$
is an oriented $S^1$-bundle over $M$
\begin{equation}
\begin{CD}
S^1 @>>> E \\
&& @V\pi VV \\
&& M \end{CD}
\end{equation}
characterized by its first Chern class $c_1(E)\in H^2(M,\ZZ)$, in the
presence of (possibly nontrivial) $H$-flux $H\in
H^3(E,\ZZ)$.\footnote{To simplify the notations we will use the same
notation for a cohomology class $[H]$, or for a representative $H$,
throughout this paper.  It should be clear which is meant from the
context.}  We will argue that the T-dual of $E$ is again an oriented
$S^1$-bundle over $M$, denoted by $\hat E$,
\footnote{Throughout this paper the notation $\hat E$ will refer to
the T-dual of the bundle $E$, and {\it not} to the dual bundle
in the usual sense.}
\begin{equation}
\begin{CD}
\hat S^1 @>>> \hat E \\
&& @V\hat \pi VV     \\
&& M \end{CD}
\end{equation}
supporting $H$-flux $\hat H\in H^3(\hat E,\ZZ)$, such that
\begin{equation} \label{eqAf}
c_1(\hat E) = \pi_* H \,,\qquad
c_1(E) = \hat \pi_* \hat H \,,
\end{equation}
where $\pi_*  : H^k(E,\ZZ) \to H^{k-1}(M,\ZZ)$, and similarly 
$\hat\pi_*$, denote the pushforward maps.\footnote{At
the level of de Rham cohomology, the pushforward maps $\pi_*$ and $\hat
\pi_*$ are simply the integrations along the $S^1$-fibers of $E$ and
$\hat E$, respectively.}

Mathematically, the reason for the duality \eqref{eqAf} can be 
understood as follows:
For an oriented
$S^k$-bundle $E$, we have a long exact sequence in cohomology
called the {\it Gysin sequence} (cf.~\cite[Prop.\ 14.33]{BT}).
In particular, for an oriented $S^1$ bundle
with first Chern class $c_1(E)=F\in H^2(M,\ZZ)$, we have
\begin{equation*}
\begin{CD}
  \ldots @>>> H^k(M,\ZZ) @>\pi^*>> H^k(E,\ZZ) @>\pi_*>> H^{k-1}(M,\ZZ)
  @> F \cup >> H^{k+1}(M,\ZZ) @>>> \ldots
\end{CD}
\end{equation*}
Consider the $k=3$ segment of this sequence.  It shows that to any
$H$-flux $H\in H^3(E,\ZZ)$ we have an associated element $\hat F
= \pi_* H \in H^2(M,\ZZ)$, and that, moreover, $F\cup \hat F=0$
in $H^4(M,\ZZ)$.   
Now, let $\hat E$ be the $S^1$-bundle associated
to $\hat F$.  Reversing the roles of $E$ and $\hat E$ in the Gysin
sequence, we see that since $F\cup \hat F=\hat F \cup F=0$,
there exists an $\hat H \in H^3(\hat E,\ZZ)$ such that $\hat \pi_*
\hat H = F$, where $\hat H$ is unique up to an element of
$\pi^* H^3(M,\ZZ)$.  The transformation $(E,H) \to (\hat E,\hat H)$,
for a particular choice of $\hat H$, is precisely what can be
identified with T-duality.  The ambiguity in $\hat H$, up to an
element in $\pi^* H^3(M,\ZZ)$, is fixed by requiring that T-duality
should act trivially on $\pi^* H^3(M,\ZZ)$, $i.e.$\ T-duality should
not affect $H$-flux which is completely supported on $M$.
Since $H$ and $\hat H$ live on different spaces, in 
order to compare them we have to pull them back to the 
correspondence space.    
The correspondence space in this more general setting is the fibered product 
$E\times_M\hat E= \{ (x,\hat x) \in E\times\hat E\ |\ \pi(x)=
\hat \pi(\hat x)\}$, which is both an $\hat S^1$-bundle over $E$,
as well as an $S^1$-bundle over $\hat E$.  

Before we continue, let us observe that in the case of a 2-dimensional
base manifold $M$, the Gysin sequence immediately gives an 
isomorphism between $H^3(E,\ZZ)$ and $H^2(M,\ZZ)$, $i.e.$ between
Dixmier-Douady classes on $E$ and line bundles on $M$.  This
correspondence
is used for example in \cite[Sect.~4.3]{Bry} to give an explicit construction 
of a $PU$-bundle (with given decomposable DD class) 
over $E$ from a linebundle 
over $M$.  As a particular concrete example,
note that $S^3$ can be considered as an $S^1$-bundle over 
$S^2$ by means of the Hopf fibration.  
By \eqref{eqAf} its 
T-dual, in the absence of $H$-flux, is $S^2\times S^1$ supported
by 1 unit of $H$-flux.  This example was studied in \cite{AABL}, but
the observation that the $H$-flux on the $S^2\times S^1$ side is 
nontrivial was apparently missed.

In order to discuss the generalization of \eqref{eqAa} we have to
choose specific representatives of the cohomology classes.  In
particular, upon choosing connections $A$ and $\hat A$, on the
$S^1$-bundles $E$ and $\hat E$, respectively, the 
isomorphism $T_*$ that generalizes \eqref{eqAa} is now given by
\begin{equation} \label{eqAh}
  \hat G = \int_{S^1} e^{ A \wedge \hat A }\ G \,,
\end{equation}
where the right hand side is a form on $E\times_M\hat E$, and 
the integration is along the $S^1$-fiber of $E$.\footnote{Strictly
speaking, the various forms entering \eqref{eqAh} and
beyond are the pullbacks
of forms on living on $E$ and $\hat E$ to $E\times_M\hat E$.}
In terms of $A$, $\hat A$, and their curvatures $F=dA$, $\hat F=d\hat
A$, we can write (see Sect.~\ref{secCa} for more details)
\begin{equation}
  H  = A\wedge \hat F -  \Omega \,,
\end{equation}
for some $\Omega\in \Omega^3(M)$, while the T-dual $\hat H$ is given
by
\begin{equation}
  \hat H  = F\wedge \hat A -  \Omega \,.
\end{equation}
Locally, we have 
$A=d\theta + \hat \pi_* \hat B$, $\hat A=d\hat\theta + \pi_* B$.
The equations \eqref{eqAf} are easily checked.  We note that
\begin{equation}
  d(A\wedge \hat A) = -H + \hat H \,,
\end{equation}
so that \eqref{eqAh} indeed maps $d_H$-closed forms to $d_{\hat
H}$-closed forms.

We recall that the RR fields $G$ are determined by 
the twisted K-theory classes $Q$ via the twisted Chern map
\cite{MM,Wib,MW,BCMMS,MS}
\begin{equation}\label{chern}
  G = ch_H(Q) {\sqrt{\widehat{A}(TE)}} \,,
\end{equation}
where $\widehat{A}$ is the A-roof genus.

The discussion above can be lifted to K-theory and,
in this more general setting, T-duality gives an isomorphism
of the twisted K-theories of $E$ and $\hat E$, descending to
an isomorphism between the twisted cohomologies of $E$ and $\hat E$,
as expressed in the following commutative diagram (see Theorem \ref{thCDa})
\begin{equation} \label{eqAg}
\begin{CD}
K^\bullet(E, H)  @>T_!>> K^{\bullet+1}(\hat E, \hat H)  \\         
      @V{ch_H}VV          @VV{ch_{\hat H}} V     \\
H^\bullet  (E, H)    @>T_*>>  H^{\bullet+1} (\hat E, \hat H)     
\end{CD}\end{equation}
Several of the constructions 
used in the definition of T-duality on twisted K-theory are adapted from
\cite{MaMS,MaMS2}.

The rationale for the normalization in (\ref{chern}) 
by ${\sqrt{\widehat{A}(TE)}}$ is fairly standard.
A special case of the cup product pairing (\ref{kcup}) 
followed by the standard index pairing of elements of K-theory 
with the Dirac operator, explains the upper horizontal arrows
in the diagram,
\begin{equation} \label{rootA}
\begin{CD}
K^\bullet(E, H)  \times K^\bullet(E, -H)  @>>> K^0(E) 
@>{\rm index}>>\ZZ \\         
      @V{ch_H}\times {ch_{-H}}VV          @VV{ch} V      @VV{||}V \\
H^\bullet(E, H)  \times H^\bullet(E, -H)    @> >>  
H^{even} (E)   @>\int_E{\widehat{A}(TE)}\wedge>>\ZZ  
\end{CD}\end{equation}
The bottom
horizontal arrows are cup product in twisted cohomology  (\ref{hcup}) 
followed by cup 
product by ${\widehat{A}(TE)} $
and by integration. By the Atiyah-Singer index theorem, the diagram 
(\ref{rootA}) 
commutes.  Therefore the normalization in (\ref{chern})
makes the pairings in twisted K-theory and 
twisted cohomology isometric.  

The twisted K-theory isomorphism is the geometric analogue of results 
of Raeburn and Rosenberg \cite{RR} who studied spaces with an $\RR$-action
in terms of crossed products of $C^*$-algebras of the type $A\rtimes_\alpha
\RR$, such that the spectrum of $A\rtimes_\alpha\RR$ is precisely the
circle bundle $E$ in the discussion above.  The isomorphism in the upper
horizontal arrow in
\eqref{eqAg} is then a direct consequence of the Connes-Thom 
isomorphism \cite{Con} of the K-theory of these crossed $C^*$-algebras.

The paper is organized as follows.
In Sect.~\ref{secB} we provide some physical intuition and motivation
for our conjectured description of T-duality although we restrict attention
to the special case in which $H$ is only nontrivial on one side of the
duality.  In Sect.~\ref{secBa}
we see how T-duality and Eqn.~(\ref{eqAf}) arise in the $E_8$ gauge 
bundle formalism of
M-theory, and in Sect.~\ref{secBb} we provide a physical derivation 
from S-duality for the case
in which $H$ is proportional to $G_3$.  Both approaches
illustrate the connection between the fibered product $E\times_M
\hat{E}$ and F-theory.  The full derivation of the isomorphism and the
corresponding maps appears in the more mathematical
Section \ref{MathSec}.  

In Sect.~\ref{ExSec} we will provide a number of
examples of this correspondence, including T-duality transverse to an
NS5-brane and T-duality of
circle bundles over Riemann surfaces which include the nilmanifolds,
Lens spaces 
and also $AdS^3\times S^3\times T^4$ with its $\Z_n$ quotients.  An example 
with torsion $H$-flux, the circle bundles over $\rpt$, will also be 
treated.  In Sect.~{\ref{rpnsec}} we consider circle bundles over $\rp^n$.  
As these
examples may be 4-dimensional or higher, we will not be able to compute 
K-groups simply by using the Atiyah-Hirzebruch spectral sequence as in the
previous section, but also we need to solve an extension problem.  However
T-duality will relate these bundles to bundles in which the extension problem 
is trivial, and so T-duality may be used to solve the extension problem in 
our original bundles and thus to calculate the twisted K-groups of circle 
bundles over $\rp^n$.

In Sect.~\ref{AnSec} we will consider the 
T-duality between $AdS^5\times S^5$ and $AdS^5\times
\CP^2\times S^1$ with $H$-flux, and its $\Z_n$ quotients \cite{BHO}.  
These are
interesting because the right hand side is not $spin$, and so one
might expect a gravitino anomaly.  However there is no gravitino
anomaly before the T-duality.  We show that in this case and in
general, as a result of the $\psi H \psi$ coupling in the
type-$\two$ supergravity
action, the nontrivial $H$-flux precisely forces the gravitino
anomalies to match before and after the T-duality.\footnote{The global 
gravitino anomaly in question is the
ill-definedness of the partition function 
that appears when an uncharged fermion is placed on a
non-$spin$ manifold, not the $(4d+2)$-dimensional chiral anomaly
discussed in, for example. Ref.~\cite{AG}.} 
We will see that the global anomalies before 
and after the T-duality agree because
they are determined by the topology of the fibered product.
In the example, this leads to an
anomaly on both sides precisely when $n$ is even.  On the other hand both 
sides are consistent when $n$ is odd, the IIB side because spacetime is 
$spin$ and the IIA side because a 9-dimensional analog of the quantum Hall 
effect in the dimensionally reduced theory means that the low energy modes of 
the gravitinos behave like bosons.  As the M-theory
lift is not $spin$, the usual formula for $G_4$ flux quantization
\cite{Wit} does not make sense, however the global gravitino
anomaly allows a new condition to be found in the torus-bundle
case.  Finally, in Sect.~\ref{GenSec}, we present some of the 
many remaining open problems.

% =========================================================================
\section{Physical motivation} \label{secB}

\subsection{T-Duality from $E_8$} \label{secBa}

The T-duality discussed in the introduction 
is a consequence of a conjecture \cite{Me} made
in the context of the $E_8$ gauge bundle formalism
\cite{Wit,DMW,Madrid,Allan,Morrison}.  In this formalism,
M-theory's 4-form fieldstrength $G_4$ is interpreted as the
characteristic class of an $E_8$ bundle $P$ over the 11$d$ bulk
$Y^{11}$.  Consider the case in which $Y^{11}$ is a 
$T^2=S^1_M\times S^1_{\twoa}$ torus bundle over
the 9-manifold $M^9$.  Dimensionally reducing out the M-theory circle 
$S^1_M$ we obtain \cite{Horava} an
$LE_8$ bundle $P\p$ over the 10-dimensional circle bundle $E$,
whose based part is characterized by a 3-form $H=\int_{S^1_M}G_4$.
$LE_8$ is the loopgroup of $E_8$.  Reducing on the other circle yields
an $LLE_8$ bundle $\hat{E}$ whose based part is characterized by a two-form
\begin{equation}
F=\int_{S^1_{\twoa}}H. \label{fh}
\end{equation}
In fact the based part of $LLE_8$ is homotopic
to the circle $S^1_{\twob}$, and so $F$ is just the curvature of a
circle bundle.
\begin{equation}
\left\{
\begin{matrix}
E_8&\to & P\cr
&&\dwn\cr
S^1_M & \to & Y^{11}\cr
&&\dwn\cr
S^1_{\twoa} & \to & E\cr
&&\dwn\cr
&&M^9
\end{matrix}
\right\}
\hsp{.2}\longrightarrow\hsp{.2}
\left\{
\begin{matrix}
LE_8&\to & P\p\cr
&&\dwn\cr
S^1_{\twoa} & \to & E\cr
&&\dwn\cr
&&M^9
\end{matrix}
\right\}
\hsp{.2}\longrightarrow\hsp{.2}
\left\{
\begin{matrix}
LLE_8\sim S^1_{\twob}&\to & \hat{E}\cr
&&\dwn\cr
&&M^9
\end{matrix}
\right\} \label{bundlez}
\end{equation}
The conjecture in Ref.~\cite{Me} is that the 
fiber of this circle bundle $S^1_{\twob}$ is
the T-dual circle which appears in $\twob$.  As desired, the first Chern
class of this bundle is precisely the $H$-flux in $\twoa$ integrated over
the fiber $S^1_{\twoa}$ as seen in Eqn.~(\ref{fh}).  
In this note we further claim that the first
Chern class of the $S^1_{\twoa}$ bundle, the spacetime on the type $\twoa$
side, is the integral over $S^1_{\twob}$ of the $H$-flux on the type $\twob$
side \eqref{eqAf}.

% =========================================================================
\subsection{T-duality from S-duality} \label{secBb}

An alternate approach to the T-duality relation \eqref{eqAf}, 
is via the F-theory \cite{Vafa} lift of this story, where
T-duality will simply be a choice of projection map.  This approach is 
similar to that of Ref.~\cite{AABL} where it was 
shown that the sigma models on 
$E$ and $\hat{E}$ may both be obtained from a sigma model on 
$E\times_M \hat{E}$ by integrating out different variables.  
Their argument, like the one in this section, only applies to the 
case in which $H$ is nonvanishing on one side of the duality, and 
the normalization is unclear.  However it may be possible to generalize 
their argument to the case in which $H$ and $\hat{H}$ are both nontrivial
(or even to higher-dimensional torii).

Recall from Eqn.~(\ref{bundlez}) that the bosonic data of M-theory is encoded
in an $LLE_8$ bundle over $M^9$.  To arrive at type $\twob$ string theory
we considered only the based part of this loop group which is
homotopic to the circle $S^1_{\twob}$, but in fact \cite{AEV} the loop
groups are free and trivially centrally extended.  Thus we find that
\begin{equation}
  \pi_1(LLE_8)=\Z^3
\end{equation}
where the three circles are $S^1_M$, $S^1_{\twoa}$\ and $S^1_{\twob}$.
These circles are all fibered over $M^9$, with Chern classes that in
type $\twoa$ we name $G_2$, $c_1(E)$\ and $c_1(\hat E)=\int_{S^1_{\twoa}} H$
respectively.  The total space of the fibered product of these three
circle bundles over $M^9$ is twelve-dimensional, and this 12$d$
perspective is called F-theory.

The total space of F-theory is an $S^1_M$ bundle over the fibered
product $E\times_M \hat E$ and also a torus bundle over $\hat E$,
the spacetime of type $\twob$.  This torus is generated by the circles
$S^1_M$ and $S^1_{\twoa}$.  Interchanging these two circles (with a
minus sign) is called S-duality in type $\twob$ and is called a 9-11 flip
in type $\twoa$.  Therefore we have the commuting diagram:

\begin{equation*} \label{correspondence}
\xymatrix @=4pc {
\twoa & &   \twob\\ 
c_1(E)=a   \ar@{<->}[rr]^{\textup{T-Duality}} &&  H=a\cup b\\
&& \\
G_2=a \ar@{<->}[uu]_{\textup{\ 9-11 Flip}} 
\ar@{<->}[rr]^{\textup{T-Duality}}  && G_3=a\cup b 
\ar@{<->}[uu]_{\textup{\ S-Duality}} 
}
\end{equation*}
relating the two $\twoa$ and two $\twob$ configurations described above.

This diagram will allow us to perform T-duality from $\twob$ to
$\twoa$ in two ways, by proceeding left directly, or by performing an
S-duality followed by a T-duality followed by a 9-11 flip.  We will
start in type $\twob$ on $M^9\times S^1_{\twob}$ with
\begin{equation}
H=a\cup b\in H^2(M^9)\otimes H^1(S^1_{\twob})
\end{equation}
and no $G_3$ flux.  Performing an S-duality leaves
$G_3=a\cup b$ and $H$ now vanishes.\footnote{Had we allowed $G_3$ flux
proportional to $H$ we could still have arranged this by performing a 
different $SL(2,\Z)$
transformation on $S^1_M$ and $S^1_{\twoa}$.}  
Now that there is no $H$
flux, we may perform a T-duality along $S^1_{\twob}$ without changing
the 10-dimensional topology.  After T-duality we find type $\twoa$ string
theory on $M^9\times S^1$ with $G_2=\int_{S^1_{\twob}}G_3=a$.  The
M-theory circle $S^1_M$ is nontrivially fibered over $M^9$ with Chern
class equal to $G_2=a$.  The 9-11 flip interchanges the M-theory
circle $S^1_M$ with the $\twoa$ circle $S^1_{\twoa}$ and so leaves $G_2=0$
and a 10-dimensional spacetime $E$ which is a $S^1_{\twoa}$ circle
bundle over $M^9$ with first Chern class
\begin{equation}
  c_1(E) = a = \int_{S^1_{\twob}}  H
\end{equation}
as desired, where $H$ is the original $H$-flux in type $\twob$.

% =========================================================================
\section{T-duality Isomorphism in Twisted K-theory and Twisted 
      Cohomology: the case of circle bundles} \label{MathSec}

\subsection{The Setup} \label{secCa}

We elaborate here on the setup in the introduction.  Suppose that $M$
is a compact connected manifold and $E$ be a principal circle bundle
over $M$ with projection map $\pi$ and $H$ a closed, integral 3-form
on $E$ having the property that $\pi_*(H)$ is a closed integral 2
-form on $M$. Then we know by the classification of circle bundles
that there is a circle bundle $\hat E$ over $M$ with projection map
$\hat \pi$ and with first Chern class $c_1(\hat E) = \pi_*(H)$.  $\hat
E$ will be referred to as the {\em T-dual} of $E$, which is not to be
confused with the dual bundle to $E$.
We define the {\em correspondence space} of $E$ and $\hat E$ to be the
fibered product $E \times_M \hat E$, since it implements T-duality in
generalized cohomology theories such as K-theory, cohomology and
their twisted analogues. Correspondence spaces also occur in other parts
of mathematical physics, such as twistor theory and noncommutative
geometry.  We have the following commutative diagram
\begin{equation} \label{correspondenceb}
\xymatrix @=8pc @ur { E \ar[d]_{\pi} & 
E\times_M  \hat E \ar[d]_{\hat p} \ar[l]^{p} \\ M & \hat E\ar[l]^{\hat \pi}}
\end{equation}

Note that the correspondence space $E \times_M \hat E $ is a circle bundle
over $E$ with first Chern class $\pi^*(c_1(\hat E))$,  and it is also a 
circle bundle over $\hat E$
with first Chern class $\hat \pi^*(c_1(E))$, by the commutativity of the 
diagram above, (\ref{correspondenceb}). If $\hat E=E$ or 
if $\hat E = M \times S^1$, 
then the correspondence space $E \times_M \hat E $ is diffeomorphic  to 
$E \times S^1$.

Let $A \in \Omega^1(E)$ and $\hat A \in \Omega^1(\hat E)$ be
connection one forms on $E$ and $\hat E$ respectively, and denote 
their curvatures in $H^2(M)$ by $F=dA$ and $\hat F=d\hat A=\pi_*H$, 
respectively.  The connections $A$ and $\hat A$ are normalized such that
$\pi_*A=1=\hat\pi_*\hat A$.
Let $H \in \Omega^3(E)$ be the given closed integral 3-form on $E$ as above.
We will now argue, as mentioned in the introduction, that there exists 
a 3-form $\Omega\in \Omega^3(M)$ such that 
\begin{equation} \label{eqCAa}
 H = A \wedge \pi^* \hat F - \pi^*\Omega  \qquad
\in  \Omega^3(E ).
\end{equation} 
Consider the Gysin sequence associated to the $S^1$-bundle $E$
(at the level of de Rham cohomology)
\begin{equation*}
\begin{CD}
  \ldots @>>> H^k(M) @>\pi^*>> H^k(E) @>\pi_*>> H^{k-1}(M)
  @> F \wedge >> H^{k+1}(M) @>>> \ldots
\end{CD}
\end{equation*}
The $k=3$ segment of the Gysin sequence shows
that $F\wedge \hat F = 0$ in $H^4(M)$.  Therefore $F\wedge\hat F = d\alpha$
with $\alpha\in\Omega^3(M)$.  Thus, $A\wedge \pi^*\hat F - \pi^*\alpha$ 
is a closed 3-form in $\Omega^3(E)$, i.e.\ an element of $H^3(E)$.  
Consider $H- (A\wedge \pi^*\hat F - \pi^*\alpha) \in H^3(E)$.   Clearly,
$\pi_*(H- (A\wedge \pi^*\hat F - \pi^*\alpha ))=0$, since $\pi_*\circ
\pi^*=0$ and $\pi_*A=1$.
Hence we conclude that $H- (A\wedge \pi^*\hat F - \pi^*\alpha ) = 
\pi^* (\beta + d\gamma)$, for some $\beta\in H^2(M)$ and 
$\gamma\in \Omega^2(M)$.  Putting $\Omega=\alpha- \beta -d\gamma$
proves \eqref{eqCAa}.  Now define $\hat H \in \Omega^3(\hat E)$ by
\begin{equation} \label{eqDc}
 \hat H = \hat \pi^* F \wedge \hat A - \hat \pi^*\Omega  \qquad
\in  \Omega^3(\hat E ).
\end{equation} 
It easily follows that $\hat H$ is closed, i.e.\ defines an element
in $H^3(\hat E)$ and that $F = c_1(E) = \hat\pi_* \hat H$ in $H^2(M)$.
I.e., to summarize, we find the relations
\begin{equation} \label{eqDd}
\pi_* H = c_1(\hat E), \quad \hat \pi{}_*\hat H 
  = c_1(E) \qquad \in H^2(M).
\end{equation}
Note that if we define
\begin{equation}
  \cB = p^*A\wedge \hat p{}^*\hat A \quad \in \Omega^2(E \times_M \hat E )
\end{equation}
then it follows that
\begin{equation}
d\cB=d(p^*A\wedge \hat p{}^*\hat A) = -p^*H + \hat p{}^*\hat H
\end{equation}
by virtue of the commutativity of the diagram \eqref{correspondenceb},
and so the pullbacks of the two $H$-fluxes are cohomologous on the 
correspondence space $E \times_M \hat E$.

% -------------------------------------------------------------------------
\subsection{T-duality in twisted cohomology}

Here we will prove T-duality in twisted cohomology. Recall that twisted
cohomology $H^\bullet(M, H)$ is by definition the ${\mathbb Z}_2$-graded 
cohomology of the
complex $(\Omega^\bullet(M), d_H)$, with differential $d_H = d -
H\wedge\;$.  Nilpotency $d_H^2=0$ follows from the fact that $H$ is a closed
3-form on $M$.  Twisted cohomology has been studied in detail in the
papers \cite{BCMMS,MS}.

The basic functorial properties of twisted cohomology are as follows: 
\begin{enumerate}
\item (Normalization) If $H=0$  then $H^\bullet (M, H) = H^\bullet (M)$.
\item (Module property) $H^\bullet (M,H)$ is a module over 
    $H^{\text{even}}(M)$.
\item (Cup product) There is a cup product homomorphism  
\begin{equation}\label{hcup}
  H^p(M, H)\otimes H^q(M,H')\to H^{p+q}(M, H + H')\,.  
\end{equation}
\item\label{hnaturality} (Naturality) If $f\colon
N\to M$ 
is a continuous map, then there is a homomorphism
\begin{equation*}
f^*:H^\bullet (M, H)\to H^\bullet (N,f^*H).
\end{equation*} 
\item\label{hpushforward} (Pushforward) If $f\colon
N\to M$ is a smooth map which is oriented, that is $TN \oplus f^*TM$
is an oriented vector bundle, then there is a homomorphism
\begin{equation*}
f_*: H^\bullet (N,f^*H) \to H^{\bullet +d} (M, H).
\end{equation*}  
where $d= \dim M - \dim N$. 
\end{enumerate}

Properties 1.\ to 4.\ were detailed in \cite{BCMMS} and \cite{MS}.
The pushforward property 5.\ is established in a manner formally
similar to the analogous property for twisted K-theory that will be
discussed below and so its proof will be omitted for sake of brevity.

We have homomorphisms
\begin{equation}
p^*: H^\bullet (E, H) \to H^{\bullet}(E \times_M \hat E, p^*H)
\end{equation}
\begin{equation}
  e^\cB : H^\bullet (E \times_M \hat E, p^*H)\to 
  H^\bullet (E \times_M \hat E, \hat p{}^*\hat H )
\end{equation}
and 
\begin{equation}
  \hat p_* : H^\bullet(E \times_M \hat E, \hat p{}^*\hat H )\to 
  H^{\bullet +1}(\hat E, \hat H)
\end{equation}
The composition of the maps 
\begin{equation}\label{cohtduality}
  T_* := \hat p_*\circ e^\cB \circ p^*: 
  H^\bullet  (E, H) \to H^{\bullet+1}(\hat E, \hat H)
\end{equation}
is called T-duality. The situation is completely symmetric
and the inverse map is  
\begin{equation}\label{-cohtduality}
  T_*^{-1} := p_* \circ e^{-\cB} \circ \hat p{}^*: 
  H^\bullet  (\hat E, \hat H) \to  H^{\bullet+1}(E, H).
\end{equation}
To summarize, we have,

\begin{theorem} In the situation described above, 
T-duality in twisted cohomology 
\begin{equation*}
T_*\ : \   H^\bullet  (E, H) \to H^{\bullet+1}(\hat E, \hat H)\,,
\end{equation*}
is an isomorphism.
\end{theorem}

On the correspondence space, we can express the isomorphism as 
\begin{equation}
  \hat G = T_*(G) = \hat p{}_* ( e^{\cB} \wedge p^* G) \,,
\end{equation}
where we notice that since $d\cB = -p^*H +\hat  p{}^*\hat H$, we have 
$d(e^\cB) = (-p^*H + \hat p{}^*\hat H) \wedge e^\cB$. So 
\begin{equation}
  d_{\hat H} \hat G = \hat p{}_* ( e^{\cB} \wedge p^* d_{H}G).
\end{equation}
It follows that $G$ is $d_H$-closed if and only if $\hat G$ is 
$d_{\hat H}$-closed.
Moreover the formula can be inverted,
\begin{equation}
  G= T_*^{-1} (\hat G)= p_* ( e^{-\cB} \wedge \hat p^* \hat G)
\end{equation}
proving the assertion.

We next describe special cases.  The
first case that we will consider is when $E$, $\hat E$ are trivial bundles
and $H=0$. This case was discussed in \cite{Hor} 
(see also \cite{Guk,Sha,OS}).  Explicitly,
$E = M\times S^1$ and $\hat E = M \times \hat S^1$, and the
connections on the respective trivial bundles are $A = d\theta$ and
$\hat A = d\hat \theta$. $\cB= d\theta\wedge d\hat\theta$ 
is the first Chern class
of the Poincar\'e line bundle $\cP$ over $S^1 \times \hat
S^1$, and $\Lambda_\cB$ is given by the exterior product with $e^\cB$,
which is equal to the Chern character of the Poincar\'e bundle
$ch(\cP)$.  In this case, the T-duality reduces to an
isomorphism,
\begin{equation}
T_* : H^\bullet(M\times S^1) \to  H^{\bullet+1}(M \times \hat S^1).
\end{equation} 

%\attn{Following example not quite correct}
Now let 
$E = M\times S^1$ be the trivial circle bundle and let
\begin{equation}
H=F\wedge d\theta \in H^2(M)\otimes H^1(S^1) \cong H^3(M\times  S^1, \ZZ)
\end{equation}
be a decomposable class on $M\times S^1$
such that $p^*H = d\hat A \wedge  d\theta \in \Omega^3(\hat E \times S^1)$.
Then by \eqref{eqDc} and \eqref{eqDd}, we 
must have $\hat p{}^*\hat H =0$ and $\cB=\hat A \wedge d\theta$
and the first Chern class $c_1(\hat E) = \pi_*H \in H^2(M, \ZZ)$. 
 
So T-duality in this case yields an isomorphism $T_* :
H^\bullet(M\times S^1, H) \to H^{\bullet+1}(\hat E)$. What is remarkable
in this case is that twisted cohomology {\em does not} have a canonical 
ring structure in general, but in this case, one can use the T-duality
isomorphism to define the {\em fusion product} on $H^\bullet (M\times
S^1, H) $.  We will generalize this as follows.

\begin{theorem} Let $X$ be a compact connected manifold, and 
let $H \in H^3(X, \ZZ)$
be a decomposable class. Then there is a fusion product on twisted
cohomology $H^\bullet (X, H)$, making it into a ring.
\end{theorem}

To prove this, we notice that a decomposable class $H$ yields a
continuous map $F = (F_1, F_2): X \to BS^1 \times S^1$, where $BS^1$
is the classifying space of $S^1$.  But we have argued before that the
T-dual of $BS^1 \times S^1$ is the total space of the universal circle
bundle $ES^1 \to BS^1$. So we can pullback the diagram
(\ref{correspondenceb}) to see that in this case, T-duality yields an
isomorphism
\begin{equation}
  T_*\  :\ H^\bullet (X, H) \to H^{\bullet+1}(\hat E)\,,
\end{equation}
that determines the fusion product on twisted cohomology. 
Here $c_1(\hat E)= k F_1^*c_1(ES^1)$,
where $[F_2]$ is $k$ times the generator. 

% ------------------------------------------------------------------------
\subsection{T-duality in twisted K-theory}

The generalization of this duality to twisted K-theory has been known
for some time \cite{RR}.  In this section we will give a geometric
description of the isomorphism along the lines of the description of
the isomorphism of twisted cohomology described above.  We will then
see that these two isomorphisms are related by the Chern map.

We first recall the definition of twisted K-theory, cf.\
\cite{Ros,BM}.  It is a well known fact that the unitary group $U$ of
an infinite dimensional Hilbert space is contractible in the norm
topology, therefore the projective unitary group $PU = U/U(1)$ is an
Eilenberg-Maclane space $K(\ZZ,2)$. This in turn implies that the
classifying space $BPU$ of principal $PU$ bundles is $K(\ZZ,3)$.  Thus
we see that $H^3(X, \ZZ) = [ X, BPU]$, where the right hand side
denotes homotopy classes of maps between the two spaces. Another well
known fact is that $PU$ is the automorphism group of the algebra of
compact operators on the Hilbert space.  So given a closed 3-form $H$
on $X$, it determines an algebra bundle $\cE_H$ up to isomorphism: a
particular choice will be assumed. This is equivalent to a particular
choice of the associated principal $PU$-bundle $P_H$ with
Dixmier-Douady invariant $[H]$.  The twisted K-theory is by
definition the K-theory of the noncommutative algebra of continuous
sections of the algebra bundle $\cE_H$. A geometric description of
objects in twisted K-theory is given in \cite{BCMMS}.

The basic properties of twisted K-theory are as follows: 
\begin{enumerate}
\item (Normalization) If $H=0$  then $K^\bullet (M, H) = K^\bullet (M)$.
\item  \label{kmodule} (Module property) $K^\bullet (M,H)$ 
is a module over $K^0(M)$.
\item (Cup product) There is a cup product homomorphism  
\begin{equation}\label{kcup}
K^p(M, H)\otimes K^q(M,H')\to K^{p+q}(M, H + H').
\end{equation}
\item\label{naturality} (Naturality) 
If $f\colon
N\to M$ is a continuous map, then there is a homomorphism
\begin{equation*}
f^!:K^\bullet (M, H)\to K^\bullet (N,f^*H).
\end{equation*}  
\item\label{pushforward} (Pushforward) 
Let $f:N\longrightarrow M$ be a smooth 
map between compact manifolds
which is K-oriented, that is 
$TN\oplus f^*TM$ is a $spin^{\mathbb C}$ vector
bundle over $N$. 
Then there is a homomorphism
\begin{equation}
  f_!\ :\  K^\bullet (N,f^*H) \to K^{\bullet +d} (M, H)\,.
\end{equation}
where $d= \dim M - \dim N$. 
\end{enumerate}

Properties 1.,\ 3.\ and  4.\ were detailed in \cite{BCMMS},
and property 2. in \cite{MS}. 
The pushforward property 5.\ will
be discussed in Sect.~\ref{pushforwardsection}, since 
it is central to our construction of T-duality. 

Using the naturality property \ref{naturality}, we have the homomorphism,
\begin{equation}\label{kpullback}
  p^! : K^j (E, H) \to K^j(E \times_M \hat E, p^*H)\,.
\end{equation}
Observe that the principal $PU$-bundles $P_{p^*H}$ and $P_{\hat p^*{\hat H}}$ 
are canonically isomorphic to $p^*P_H$
and  $\hat p^*P_{\hat H}$, respectively. Since $-p^*H + \hat p^*\hat H 
= d\cB$, we conclude that 
$P_{p^*H}$ and $P_{\hat p^*{\hat H}}$ are isomorphic.

We digress to discuss automorphisms of twisted K-theory. First 
recall that tensoring by any line bundle on $E$ is an automorphism of 
K-theory, $K^\bullet(E)$ (for example, tensoring by the 
Poincar\'e line bundle 
on the torus). By the module property 2. of twisted K-theory, we see that 
tensoring by any line bundle on $E$ is also an automorphism of twisted 
K-theory, $K^\bullet(E, H)$. However, any line bundle on $P_H$ also 
gives rise to an automorphism of twisted K-theory as explained next.
The first fact that is needed is that stably equivalent bundle gerbes
(i.e. tensoring by a trivial gerbe)  define
the same twisted K-theory, cf. \cite{BCMMS}. The next fact is 
that any line bundle on $P_H$  determines a trivial bundle
gerbe, which when tensored with the lifting bundle gerbe of $P_H$, 
defines a bundle gerbe that is stably equivalent to the lifting bundle 
gerbe of $P_H$. 

Next we recall the homomorphism $\psi: PU \times PU \to PU$ that is
not the group multiplication, but is defined as follows. Choose an
isomorphism of the infinite dimensional Hilbert spaces $\phi: H\otimes
H \to H$. This induces an isomorphism $\phi : B(H) \times B(H) \to
B(H)$ defined by $\phi(A, B) (v) = \phi A\otimes B
(\phi^{-1}(v))$. This restricts to a homomorphism $\phi : U \times U
\to U$, where $U$ denotes the unitary operators, such that $\phi(U(1)
\times U(1)) \subset U(1)$.  Therefore we get the induced homomorphism
on the quotient $\psi: PU \times PU \to PU$.

Let $\lambda : P_H \to E $ be the principal $PU$-bundle over $E$ with
curving $f$ and 3-curvature $H$. That is $df = \lambda^*H$.  We also
make similar choices $\hat \lambda : P_{-\hat H} \to \hat E $ with curving
$-\hat f$ and 3-curvature $-\hat H$ satisfying $d(-\hat f) = -\hat
\lambda^*\hat H$.  Then on the correspondence space $E\times_M \hat
E$, we can form the trivial bundle gerbe $\tilde \lambda : P= (p^*P_H
\times {\hat p}^*P_{-\hat H}) \times_\psi PU \to E\times_M \hat E$
which has curving $f - \hat f$ and 3-curvature $H - \hat H$ (which is
equal to $-d\cB$).  We have simplified the notation by omitting some
of the pullback maps, since it is clear on which space the
differential forms live.  
Since by definition, $\pi_* A = 1$ and $\hat\pi_* \hat A = 1$, 
we see that $\cB$ is an integral 2-form. Since
$H$ and $\hat H$ are integral 3-forms, we can choose 
$f$ and $\hat f$ to be integral 2-forms.
Observe that the following identity holds,
\begin{equation}
  d(f-\hat f) = \tilde \lambda^*(H-\hat H) = d (-\tilde \lambda^* \cB)\,.
\end{equation} 
It follows that $\tilde \lambda^*\cB + f - \hat f \in
\Omega^2(P)$ is a closed 2-form on the trivial gerbe $P$
that has
integral periods, and therefore determines a line bundle $\cL\to P$
over the trivial bundle gerbe $P$, with curvature 
$ \cB + f - \hat f $ and first Chern class 
$c_1(\cL) = [\cB + f - \hat f ] $. By the discussion
above, tensoring by the trivial bundle gerbe determined by this line
bundle $\cL$ induces the following isomorphism in twisted K-theory,
\begin{equation} \label{kuntwist}
  \Lambda_\cB : K^j(E \times_M \hat E, p^*H)\to K^j(E \times_M \hat E, 
  \hat p{}^*\hat H ) \,.
\end{equation}
Using the pushforward property \ref{pushforward}, we have a homomorphism,
\begin{equation}
  \hat p_! : K^j(E \times_M \hat E, \hat p{}^*\hat H )\to 
  K^{j+1}(\hat E, \hat H) \,.
\end{equation}
The composition of the maps 
\begin{equation}\label{tduality}
  T_!:= \hat p_! \circ \Lambda_\cB \circ p^! : 
  K^j (E, H) \to  K^{j+1}(\hat E, \hat H)
\end{equation}
is the T-duality in twisted K-theory. The situation is 
completely symmetric
and the inverse map is  
\begin{equation}\label{-tduality}
  T_!^{-1} := p_! \circ \Lambda_{-\cB} \circ \hat p{}^! : 
  K^j (\hat E, \hat H) \longrightarrow  K^{j+1}(E, H)\,.
\end{equation}
To summarize, we have

\begin{theorem} In the situation described above, 
T-duality in twisted K-theory,  
\begin{equation*}
  T_!\  :\ K^\bullet (E, H) \longrightarrow  K^{\bullet+1}(\hat E, \hat H)
\end{equation*} 
is an isomorphism.
\end{theorem}

The special cases discussed above in the context of twisted cohomology
are virtually identical in the case of twisted K-theory.  In
particular, in the decomposable case we find a ring structure
(cf.\ \cite{FHT}).

\begin{theorem} 
Let $X$ be a compact connected manifold, and let $H\in H^3(X, \ZZ)$
be a decomposable class. Then there is a fusion product on twisted K-theory
$K^j (X, H)$,  making it into a ring.
\end{theorem}

% -------------------------------------------------------------------------
\subsection{The pushforward map}\label{pushforwardsection}

In this section we define the pushforward of a K-oriented map in
twisted K-theory, $i.e.$ property \ref{pushforward}. We shall see in
this section that this is essentially the topological index in
\cite{MaMS,MaMS2}, and we will follow the construction given there.
\begin{equation} \label{shriek}
\xymatrix @=4pc { 
&  N( E\times \RR^{2N}/Z) \cong U \ar@{^{(}->}[dr]^{i_1} & \\
Z \ar@{^{(}->}[ur]^{{j_1}_!} \ar[dd]_p \ar@{^{(}->}[rr]^j & & 
  E\times \RR^{2N} \ar[dd]_{p_1}\\
 & & \\
E \ar[rr]^= \ar@{^{(}->}[uurr]^i & & E 
}
\end{equation}

For the discussion below, we will make use of the commutative diagram
above, which we now explain. Given a fiber bundle $p : Z\to E $ where
the projection map $p$ is K-oriented, there is an embedding $i: Z
\hookrightarrow E \times \RR^{2N}$ that commutes with the projection
map $p$, cf.~\cite{AS}.  Let $i : E \hookrightarrow E \times \RR^{2N}$
be the zero section embedding and $p_1 : E\times \RR^{2N} \to E$ the
projection map to the first factor. Now the total space $Z$ embedds as
the zero section of the normal bundle to the embedding $j$, $i.e.$
$j_1 : Z \hookrightarrow N(E\times \RR^{2N}/Z)$. The normal bundle
$N(E\times \RR^{2N}/Z)$ is diffeomorphic to a tubular neighborhood $U$
of the image of the correspondence space in $E \times
\RR^{2N}$. Finally, $i_1 : U \hookrightarrow E \times \RR^{2N}$ is the
inclusion map.

\begin{lemma}\label{Bott} There is a
canonical
isomorphism
\begin{equation*}
  i_! \ :\ K^\bullet (E, H)\cong K_c^\bullet (E\times
  \R^{2N}, p_1^*H)
\end{equation*}
that is determined by Bott
periodicity.
\end{lemma}
\begin{proof} Recall that 
$K_c^\bullet (E\times \R^{2N}, p_1^*H) = 
K_\bullet (C_0(E\times \R^{2N},\cE_{p_1^*H}))$.  
Now there is a canonical isomorphism
$\cE_{p_1^*H}
\cong p_1^*\cE_{H}$, which induces a canonical
isomorphism $C_0(E\times \R^{2N},
\cE_{p_1^*H}) \cong C(E,\cE_{H})
\widehat\otimes
C_0(\R^{2N})$.  Thus, $K_c^\bullet (E\times \R^{2N},
p_1^*H) \cong
K_\bullet (C(E,\cE_{H}) \otimes C_0(\R^{2N}))$.
Bott periodicity asserts that, $K_\bullet (C(E,\cE_{H}) \otimes
C_0(\R^{2N})) \cong K^\bullet (E, H),$ proving the
lemma.
\end{proof}
Our goal is to next define $j_!: K^\bullet (Z,
p^*H) \longrightarrow K_c^\bullet (E\times \RR^{2N}, p_1^*H )$.  To do
this, we first consider
\begin{equation}
\begin{aligned}
  {j_1}_!\ :\  
  K^\bullet (Z, p^*H) &\longrightarrow K_c^\bullet (N(E\times \RR^{2N}/Z), 
  \pi_1^*H ),\\
  \xi &\longmapsto \pi_1^*\xi \otimes (\pi_1^* {S^+}, 
  \pi_1^*{S^-}, c(v))
\end{aligned}
\end{equation}
where $\pi_1: N(E\times \RR^{2N}/Z) \to Z$ is the projection and
$(\pi^* {S^+}, \pi^*{S^-}, c(v))$ is the usual Thom class
of the complex vector bundle $N(E\times \RR^{2N}/Z)$.  On the 
the right hand side the
we have used the module property \ref{kmodule}. 
The {Thom isomorphism}
in this context, cf\@.\ \cite{MaMS2}, asserts that $\; {j_1}_!\;$ is
an isomorphism.
Now, $N(E\times \RR^{2N}/Z) $ is diffeomorphic to a tubular neighborhood 
$U$ of the image of
$Z$ in $E\times \RR^{2N}$: let $\Phi: U 
\longrightarrow N(E\times \RR^{2N}/Z)$ 
denote this diffeomorphism. We have
\begin{equation*}
  \Phi^!\circ {j_1}_! \ : \ 
  K_c^\bullet (Z, p^*H)\longrightarrow K_c^\bullet (U, \Phi^*\pi_1^*H)\,.
\end{equation*}
The inclusion of the open set $U$ in $E \times R^{2N}$ induces a map
$K_c^\bullet (U, \Phi^*\pi_1^*H)\longrightarrow K_c^\bullet (E\times R^{2N},
p_{1}^{*}H)$. The composition of these maps defines the Gysin
map. In particular we get the Gysin map in twisted K-theory,
\begin{equation*}
  j_! \ :\   K^\bullet (Z, p^*H) \longrightarrow K_c^\bullet ( E \times
  \R^{2N}, p_1^*H)\,,
\end{equation*}
where $ j_!  = i_1 \circ \Phi^!\circ {j_1}_! $.
Now define the \emph{pushforward}
\begin{equation*}
  p_! =i_!^{-1}\circ j_! : K_c^\bullet (Z, p^*H)
  \longrightarrow K^\bullet (E, H)\,,
\end{equation*}
where we apply Lemma~\ref{Bott} to see that the inverse $j_!^{-1}$ 
exists.

This defines the pushforward for submersions and immersions.  The
general case can be deduced in the standard manner.  Let $f : N \to M$
be a smooth map that is K-oriented. Then $f$ can be canonically
factorized into an embedding followed by a submersion as follows.
Consider the graph embedding $i_f: N \hookrightarrow N\times M$
defined by $i_f(n) = (n, f(n))$, which is K-oriented since $f$ is
K-oriented, and the submersion $p_2 : N\times M \to M$, which is
also K-oriented for the same reasons.  Then we already know how to
define the homomorphisms
\begin{equation*}
  {i_f}_! : K^\bullet (N, f^*H) \to K^\bullet (N \times M, p_1^*H) \,,
\end{equation*}
and also 
\begin{equation*}
  {p_2}_! : K^\bullet (N \times M, p_1^*H) \to K^\bullet (M, H)\,.
\end{equation*}
Define the pushforward of a general K-oriented map as
\begin{equation}  \label{kpushforward}
  f_! =  {p_2}_! \circ {i_f}_!  \,.
\end{equation}

% -----------------------------------------------------------------------
\subsection{T-duality and twisted Grothendieck-Riemann-Roch formulae}

We will first recall the twisted Chern character $ch_H :
K^\bullet(E,H) \to H^\bullet(E,H)$ and then compute the twisted Chern
character of the T-dual of an element in twisted K-theory.  
Since for dimension reasons 
${\rm Todd}(T^{vert}E) = 1= {\rm Todd}(T^{vert}\hat E)$, 
this yields the following,

\begin{theorem} \label{thCDa}
In the notation of section 3, 
there is a commutative diagram,
\begin{equation} \label{grr}
\begin{CD}K^\bullet(E, H)  @>T_!>> K^{\bullet+1}(\hat E, \hat H)  \\         
      @V{ch_H}VV          @VV{ch_{\hat H}} V     \\
H^\bullet  (E, H)    @>T_*>>  H^{\bullet+1} (\hat E, \hat H)     
\end{CD}\end{equation}
The Grothendieck-Riemann-Roch formula in this context expresses 
this commutativity,
\begin{equation}\label{grr2}
ch_{\hat H}(T_!(Q)) = T_*(ch_H(Q)) 
\end{equation}
for all $Q\in K^\bullet(E,H)$. 
\end{theorem}

Eqn.\ (\ref{grr2}) can be re-expressed as 
\begin{equation}\label{grr3}
  ch_{\hat H}(T_!(Q)) = \hat p_*(e^\cB \wedge ch_H(Q)).
\end{equation}

We begin by recalling that in \cite{BCMMS} a homomorphism 
$ch_{H}\colon {K}^0
(E,H)\to H^{even}(E, H)$ was constructed with the following
properties: \\
1) $ch_{H}$ is natural with respect
to pullbacks, \\
2) $ch_{H}$ respects the ${K}^0(E)$-module
structure of ${K}^0(E, H)$,  \\
3) $ch_{H}$ reduces
to the ordinary Chern character in the untwisted case
when $H = 0$.  \\
It was proposed that $ch_{H}$ was the
Chern character for  twisted K-theory.  We
give a heuristic construction of $ch_{H}$ here,
referring to \cite{BCMMS} and \cite{MS} for details.

Let
$\lambda: P_H\to E$ be a principal $PU$ bundle 
with given gerbe connection, and curving to be explained below. 
Let $\cE_i \to P$ be $U_{\rm tr}$-modules for the lifting
bundle gerbe $L\to {P_H}^{[2]}$, where 
 $U_{\rm tr}$ denotes the unitary operators of the form
identity plus trace class - then $[\cE_1]-[\cE_0] \in K^0(E,H)$.  
That is, there is an action
of $L$ on $\cE_i$ via an isomorphism $\psi\colon
\pi_1^*\cE_i\otimes L\to \pi_2^*\cE_i$.  
We suppose that
$L$ comes equipped with a bundle gerbe connection
$\nabla_L$ and a choice of curving $f$ such that
the associated $3$-curvature is $H$, a closed,
integral $3$-form on $E$ representing the image,
in real cohomology, of the Dixmier-Douady class of $P_H$.  
Since
the ordinary Chern character $ch$ is multiplicative,
we have
\begin{equation}
\label{character equation}
\pi_1^*(ch(\cE_1) - ch(\cE_0))ch(L) = \pi_2^*(ch(\cE_1) - ch(\cE_0)).
\end{equation}
It turns out that this equation holds
on the level of differential forms.  Then $ch(L)$ is represented
by the curvature $2$-form $F_L$ of the
bundle gerbe connection $\nabla_L$ on $L$.  A choice
of a curving for $\nabla_L$ is a $2$-form $f$ on $P_H$
such that $F_L = \delta(f) = \pi_1^*f -
\pi_2^*f$ and $f$ has the property that $df = \lambda^*H$.
 It follows that $ch(L)$ is represented
by $\exp(F_L) = \exp(\pi_1^*f - \pi_2^*f) =
\exp(-\pi_2^*f)\exp(\pi_1^*f)$.  Therefore we can rearrange
the equation~(\ref{character equation}) above to get
\begin{equation}
\label{differential character equation}
\pi_1^*\exp(f)(ch(\cE_1) - ch(\cE_0)) = \pi_2^*\exp(f)
(ch(\cE_1) - ch(\cE_0)).
\end{equation}
Since we are assuming that this equation 
\eqref{differential character equation} holds at the
level of differential forms, this implies that the differential form
$\exp(f)(ch(\cE_1) - ch(\cE_0)) $ descends to to a differential 
form on $E$ which
is clearly closed with respect to the twisted
differential $d-H$, and is the Chern-Weil representative of the 
twisted Chern character.
That is, $\lambda^*ch_H(\cE_1-\cE_0) = \exp(f)(ch(\cE_1) - ch(\cE_0)). $
We will use the simplified notation, 
\begin{equation}\label{twistch}
\lambda^*ch_H(Q) = e^{f}ch(Q), \qquad Q\in K^0(E, H).
\end{equation}
In section 5, \cite{MS}, a similar formula was obtained for the 
odd twisted Chern character,
\begin{equation}\label{twistch2}
\lambda^*ch_H(Q) = e^{f}ch(Q), \qquad Q\in K^1(E, H).
\end{equation}

We next study the Grothendieck-Riemann-Roch formula
in twisted K-theory, following the computation of the Chern
character of the topological index in \cite{MaMS,MaMS2}. 
Let $\tau: \cQ \longrightarrow E$ be a $spin^\CC$ vector bundle over $E$ and 
$i: E\longrightarrow \cQ$ the zero section embedding.
Let $P_H$ be the principal $PU$-bundle over $E$: then for 
$\xi \in K^\bullet(E, H)$, we compute,
$$
\begin{array}{lcl}
ch_{\tau^*H}(i_! \xi) &=& ch_{\tau^*H}(i_! 1 \otimes \tau^*\xi)\\[+7pt]
		 &=& ch(i_! 1) \cup ch_{\tau^*H}(\pi^* \xi),
\end{array}
$$ 
where we have used the fact that the Chern character
respects the $K^0(E)$-module structure.
The standard Riemann-Roch formula asserts that
\begin{equation*}
   ch(i_! 1) = i_*{\rm Todd}(\cQ)^{-1} = i_*1 \cup 
  \tau^*{\rm Todd}(\cQ)^{-1}\,.
\end{equation*}
Therefore we obtain the following Riemann-Roch formula for linear 
embeddings in twisted
K-theory,
\begin{equation}\label{RR}
  ch_{\tau^*H}(i_! \xi) = i_*\left\{{\rm Todd}(\cQ)^{-1} \cup ch_H( \xi)
  \right\}\,.
\end{equation}

We will refer to the commutative diagram (\ref{shriek}) in what 
follows. 
Now $p_! = i_!^{-1} \circ j_! $, therefore for 
$\Xi \in K^\bullet(Z, p^*H)$, 
\begin{equation*}
  ch_H (p_!\Xi ) = ch_H (i_!^{-1} \circ j_!  \Xi) \,.
\end{equation*}
By the Riemann-Roch formula for linear embeddings in twisted 
K-theory cf.~\eqref{RR},
\begin{equation*}
  ch_{p_1^* H}(i_!\Xi) =  i_*ch_H(\Xi) \,,
\end{equation*}
since $p_1: E\times \RR^{2N} \longrightarrow E$ is a trivial bundle.
Since
${p_1}_* i_* 1 = (-1)^n$, it follows that
for $\ell \in K_c^\bullet (E \times \RR^{2N}, p_1^*H)$, 
one has
\begin{equation*}
  ch_H(i_!^{-1} \ell) = (-1)^n{p_1}_*ch_{p_1^*H}( \ell) \,.
\end{equation*}
Therefore
\begin{equation} \label{a1}
  ch_H (i_!^{-1} \circ j_! \, \Xi)
  =  (-1)^n p{_1}_*ch_{p_1^*H}( j_! \, \Xi ) \,.
\end{equation}
By the Riemann-Roch formula for linear embeddings in twisted K-theory 
\eqref{RR},
\begin{equation} \label{a2}
  ch_{p_1^*H}(j_!  \Xi) = j_*\left\{{\rm Todd}(N)^{-1}
  \cup ch_{p^*H}( \Xi)\right\} \,,
\end{equation}
where $N = N(E \times \RR^{2N}/Z)$ is the complex normal 
bundle to the embedding
$j : Z \longrightarrow E \times \RR^{2N}$.
Therefore ${\rm Todd}(N)^{-1} = {\rm Todd}(T(Z/E))$ and
\eqref{a2} becomes
\begin{equation*}
  ch_{p_1^*H}( j_!  \Xi) =j_*\left\{{\rm Todd}(T(Z/E))
  \cup ch_{p^*H}( \Xi)\right\}  \,.
\end{equation*}
Therefore \eqref{a1} becomes
\begin{equation}
\begin{array}{lcl}
  ch_H (i_!^{-1} \circ  j_!  \Xi ) &= &
  (-1)^n {p_1}_*j_*\left\{{\rm Todd}(T(Z/E))  \cup
  ch_{p^*H}( \Xi)\right\}\\[+7pt]
  & = & (-1)^n p_*\left\{{\rm Todd}(T(Z/E))  \cup
  ch_{p^*H}( \Xi)\right\}
\end{array}
\end{equation}
since $p_* = {p_1}_* j_*$.  Therefore
\begin{equation}\label{grrsubmersion}
  ch_H (p_! \Xi) = (-1)^n p_* \left\{{\rm
  Todd}(T(Z/E)) \cup ch_{p^*H}( \Xi)\right\}\,,
\end{equation}
proving the Grothendieck-Riemann-Roch for K-oriented submersions. For
a general K-oriented smooth map $f : N \to M$, we have seen that it
can be factorized as $ f = p_2 \circ i_f$, where $i_f : N \to N \times
M$ is the graph embedding, and $p_2 : N \times M \to M$ is the
submersion given by projection onto the second factor.  Since $ f_! =
{p_2}_! \circ { i_f}_!$ andusing the fact that we have obtained the
Grothendieck-Riemann-Roch theorem for immersions and submersions in
twisted K-theory, we can deduce it in the general case to get,
\begin{equation}\label{grrgeneral}
  ch_H (f_! \Xi) = (-1)^n f_* \left\{{\rm
  Todd}(TN/f^*TM) \cup
  ch_{f^*H}( \Xi)\right\}\,.
\end{equation}

The pulbacks and tensor products commute with the Chern map by the
functoriality of the characteristic class, and so we need only verify
that the pushforward commutes.  In this case $N$ is the correspondence
space and $M$ is $\hat E$ and so $TN/f^*TM$ is one-dimensional, too
small to have a nontrivial Todd class.  Equation (\ref{grrgeneral})
then reduces to (\ref{grr2}) up to a sign which may be absorbed into
the definition of the K-theory pushforward map.  We can apply this now
to the commutative diagram
\eqref{correspondenceb} to deduce 
the formula \eqref{grr2} in the Theorem  \ref{thCDa}.

Using \eqref{twistch},  \eqref{twistch2} and simplifying the notation, 
we compute,
\begin{equation}
\begin{array}{lcl}
ch_{\hat H}(T_!(Q)) &=& ch_{\hat H}({\hat p}_! (\cL \otimes Q))\\[+7pt]
            & = & {\hat p}_*(ch_{\hat H}(\cL \otimes Q))\\[+7pt]
        & = & {\hat p}_*(e^{\hat f} ch(\cL \otimes Q))\\[+7pt]
 & = & {\hat p}_*(e^{\hat f} e^{c_1(\cL)} ch(Q))\\[+7pt]
 & = & {\hat p}_*(e^{\hat f} e^{\cB + f - \hat f} ch(Q))\\[+7pt]
& = & {\hat p}_*(e^{\cB} e^{  f } ch(Q))\\[+7pt]
& = & {\hat p}_*(e^{\cB} ch_H(Q)) = T_*(ch_H(Q)),
\end{array}
\end{equation}
proving Theorem \ref{thCDa}.

It is possible to refine Theorem \ref{thCDa} to an equality on the
level of differential forms, using the method in \cite{MQ} - this will
be done elsewhere.

% =========================================================================
\section{3-Dimensional Examples} \label{ExSec}

\subsection{Circle bundles over the 2-torus} \label{nil}

Our first example is a slight generalization of a well-known example
related to the Scherk-Schwarz
compactification of string theory on $M^7\times T^3$
(see, e.g., \cite{GLMW,KSTT}).
Consider the 3-dimensional manifold $E$, a so-called nilmanifold,
with metric 
\begin{equation}
  g = dx^2 + dy^2 + (dz + jx\, dy)^2 \,,
\end{equation}
and $H$-flux
\begin{equation}
  H = k \, dx\wedge dy\wedge dz \,,
\end{equation}
where the coordinates $(x,y,z)$ are subject to the identifications
\begin{equation}
  (x,y,z) \sim (x,y+1,z) \sim (x,y,z+1) \sim (x+1,y,z-jy) \,.
\end{equation}
We can think of $E$ as an $S^1$-bundle over $T^2 = \{(x,y)\}$ by
\begin{equation}
  (x,y,z) \sim (x+1,y,z-jy) \,.
\end{equation}
The $S^1$-bundle has a connection $A = dz + jx\, dy$, with first 
Chern class $c_1(E) = dA = j\, dx\wedge dy$, and 
\begin{equation}
  \int_E H = k \,,\qquad \int_M c_1(E) = j \,.
\end{equation}
Let $\kappa = \partial/\partial z$ denote the Killing vector field 
associated with the circle action, $i.e.$
$\cL_\kappa g = 0 = \cL_\kappa H$.  Consider the coordinate patch
$x\in (0,1)$.  We choose a gauge in which 
\begin{equation}
  B = k x \, dy\wedge dz \,,
\end{equation}
so that $\cL_\kappa B=0$, and we can apply the Buscher rules 
\cite{Bus} (see, e.g.,
App.~A in \cite{KSTT} for a concise summary of these rules).
We find a T-dual metric and B-field given by
\begin{align}
  \hat g & = dx^2 + dy^2 + (d\hat z + kx\, dy)^2 \,,\nonumber \\
  \hat B & = j x \, dy\wedge d\hat z\,.
\end{align}
\textit{I.e.}, the T-dual corresponds again to an $S^1$-bundle over $T^2$,
this time with $H$-flux related to the initial configuration by the
interchange $j\leftrightarrow k$, in accordance with
Eqn.~\eqref{eqAf}.  Note, moreover, that
\begin{equation}
  A\wedge \hat A = dz\wedge d\hat z - 
  kx \, dy\wedge dz + jx\, dy\wedge d\hat z 
  = dz\wedge d\hat z -B + \hat B \,,
\end{equation}
so that locally Eqn.~\eqref{eqAh} does indeed agree with 
Eqn.~\eqref{eqAa}.
For a discussion of the isomorphism of K-theories we refer
to the next section, where the more general case of circle bundles over
a Riemann surface is discussed.

Note that this particular example clearly illustrates the possible
obstruction to T-dualizing over a two-torus (cf.\ the discussion in
\cite{KSTT}).  Upon starting with a three-torus (the case $j=0$ in the 
above), with $k$ units of $H$-flux and three commuting circle actions,
T-dualizing over one circle leaves us with a circle bundle (the
nilmanifold) with only one (global) $S^1$-action left, the circle
action on the dual $S^1$.

% --------------------------------------------------------------------
\subsection{Circle Bundles on a Riemann Surface}

In this section we will find the twisted K-groups of circle bundles over
2-manifolds and their T-duals and show that $K^0$ of each space
is related to $K^1$ of its dual.  This class of examples will be seen
to include the familiar examples of NS5-branes, 3-dimensional Lens
spaces and nilmanifolds.  The K-groups in the examples of this section 
(but not the next) will be
uniquely determined by the Atiyah-Hirzebruch spectral sequence
\cite{Ros,MMS}.  In fact it will suffice to consider only the
first differential
\begin{equation}
  d_3=Sq^3+H
\end{equation}
of the sequence.  Furthermore the $Sq^3$ term will be trivial, although it
would be interesting to test this correspondence in an example in
which the $Sq^3$ term is nontrivial.
Thus the K-classes will consist of cohomology classes whose cup
product with the NS fieldstrength $H$ vanishes quotiented by those
classes that are themselves cup products of classes by $H$.
Explicitly, if $H^{even}(E,\ZZ)$ and $H^{odd}(E,\ZZ)$ are the even and odd
cohomology classes of the manifold $E$ with integer coefficients, then
the twisted K-groups are
\begin{equation}
K^0(E,H)=\frac{\textup{ker}(H\cup:H^{even}\rightarrow H^{odd})}
{H\cup H^{odd}(E,\ZZ)}\,,\quad
K^1(E,H)=\frac{\textup{ker}(H\cup:H^{odd}\rightarrow H^{even})}
{H\cup H^{even}(E,\ZZ)}\,.
\end{equation}
More precisely, this procedure only yields the associated graded
algebras of the twisted K-theory, to find the actual K-groups from
these one must in general solve an extension problem.  That is to say,
torsion classes in $H^p(E,H)$ may mix with classes in $H^{p+2}$,
yielding the wrong answer.  However $H^p$ only has torsion classes for
$p\geq 2$ and $H^{p+2}$ is only nontrivial for manifolds of dimension
$d\geq p+2$.  Thus the associated graded algebras only differ from the
K-groups for manifolds of dimension $d\geq p+2\geq 4$.  In this
section we will consider only 3-dimensional examples and so will not
need to concern ourselves with the extension problem.  In the next
section we will.

Circle bundles $E$ over a manifold $M$ are entirely classified 
by their first Chern class
\begin{equation}
  c_1(E) = F \in H^2(M,\ZZ) \,,
\end{equation}
where $F$ is the curvature of the bundle and $H^2(M,\ZZ)$ 
is the manifold´s
second cohomology group with integer coefficients.%  
\footnote{Factors of $2\pi$ will be systematically absorbed 
into curvatures to make all quantities integral.}  In the case of an
orientable 2-manifold, like the 2-sphere or a more general genus $g$
Riemann surface, $H^2(M,\ZZ)=\ZZ$ and so topologically circle bundles are
classified by an integer $j$.

If the circle bundle is the trivial bundle $j=0$, then the cohomology
of the total space $E$ of the bundle is given by the K\"unneth formula
\begin{equation}
H^0(E,\ZZ)=\Z\sp
H^1(E,\ZZ)=\Z^{2g+1}\sp
H^2(E,\ZZ)=\Z^{2g+1}\sp
H^3(E,\ZZ)=\Z\,.
\end{equation}
A quick application of the Meyer-Vietoris sequence shows that if the
Chern class is equal to $j\neq 0$ then the cohomology of $E$ is
\begin{equation}
H^0(E,\ZZ)=\Z\sp
H^1(E,\ZZ)=\Z^{2g}\sp
H^2(E,\ZZ)=\Z^{2g}\oplus\Z_j\sp
H^3(E,\ZZ)=\Z\,.
\end{equation}
The $\Z^{2g}$'s will not play any important role in what follows, and
so the reader may choose to ignore them and consider the only the
2-sphere case, $g=0$.

The $H$-flux inhabits $H^3(E,\ZZ)=\Z$ and so the possible flux is
classified by another integer $k$.    We will always choose a 
basis for $H^2$ and $H^3$ such that $j$ and $k$ are nonnegative. 
The cup product with an element of
$H^3$ increases the dimension of a cocycle by $3$, so it is only
nontrivial on $0$-cocycles, which it maps to $3$-cocycles:
$H^0\rightarrow kH^3$.  If $k=0$ then $H=0$ and so everything is in
the kernel of $d_3=H\cup$.  The image of $H\cup$ in this case is
trivial, and so the untwisted K-theory is simply the cohomology
\begin{eqnarray}
  K^0(E,{H=0})&=&H^0(E,\ZZ)\oplus H^2(E,\ZZ)=\left\{
  \begin{matrix} \Z^{2g+2}\hsp{.4}&\hsp{.2}\textup{if $j=0$}\,,\cr 
  \Z^{2g+1}\oplus\Z_j&\hsp{.2}\textup{if $j\neq 0$}\,,
  \end{matrix}
  \right.\nonumber\\
  K^1(E,{H=0})&=&H^1(E,\ZZ)\oplus H^3(E,\ZZ)=\left\{
  \begin{matrix} \Z^{2g+2}\hsp{.4}&\hsp{.2}\textup{if $j=0$}\,,\cr 
  \Z^{2g+1}\hsp{.4}&\hsp{.2}\textup{if $j\neq 0$}\,.
  \end{matrix}
  \right.
\end{eqnarray}
If $k\neq 0$ then the kernel of $H\cup$ consists of all cocycles of
dimension greater than 0.  The image consists of all 3-cocycles that
are multiples of $k$, that is, the image is $kH^3(E,\ZZ)=k\Z$.  The
quotient of the kernel by the image yields the K-groups
\begin{eqnarray}
  K^0(E,{H=k})&=&H^2(E,\ZZ)=\left\{
  \begin{matrix} \Z^{2g+1}\hsp{.2}&\hsp{.2}\textup{if $j=0$}\,,\cr 
  \Z^{2g}\oplus\Z_j&\hsp{.2}\textup{if $j\neq 0$}\,,
  \end{matrix}
  \right.\nonumber \\
  K^1(E,{H=k})&=&H^1(E,\ZZ)\oplus H^3(E,\ZZ)/kH^3(E,\ZZ)=\left\{
  \begin{matrix} \Z^{2g+1}\oplus\Z_k\hsp{.2}&\hsp{.2}\textup{if $j=0$}\,,\cr 
  \Z^{2g}\oplus\Z_k\hsp{.2}&\hsp{.2}\textup{if $j\neq 0$}\,.
  \end{matrix}
  \right.
\end{eqnarray}
According to Eqn.~(\ref{eqAf}) T-duality is the interchange of $j$ and
$k$.  In every case above this results in the twisted K-groups
$K^0(E,H)$ and $K^1(E,H)$ being interchanged, which corresponds to the
fact that RR fieldstrengths are classified by $K^0(E,H)$ in type
$\twoa$ string theory and by $K^1(E,H)$ in $\twob$.  This means that
one can find the new RR fieldstrengths from the old ones by applying
the isomorphism between the two K-groups.\footnote{The general
prescription for computing the dual fieldstrengths is given in
Sect.~\ref{MathSec}.}  In this example it is quite straightforward,
one simply interchanges the $\Z^{2g}$ between $H^1$ and $H^2$ and the
rest of the cohomology groups are swapped $H^0\leftrightarrow H^1,\
H^2\leftrightarrow H^3$.

% ----------------------------------------------------------------------------
\subsection{Comparison with the Literature}

Several subcases of this class of examples have been studied in the
literature.  For example, consider type $\two$ string theory on
$\R^9\times S^1$ with a stack of $k$ NS5-branes at the same point in
a transverse $\R^3\times S^1$.  Consider a 2-sphere $S^2\subset
\R^3$ such that $S^2\times S^1$ links the stack once.  The
generalization to an arbitrary Riemann surface is straightforward.
The integral of $H$ over $S^2\times S^1$ follows from Gauss' law
\begin{equation}
  \int_{S^2\times S^1} H=k\,.
\end{equation}
The circle is trivially fibered over $S^2$ and so, in the above
notation, the first Chern class $j$ vanishes.

T-duality interchanges $j$ and $k$, which means that the T-dual
configuration has no $H$-flux, so that the NS5-branes have disappeared.
Instead the circle bundle is now nontrivially fibered, with a first
Chern class of $k$ over each Riemann surface that links (once) the
place where the stack was.  This configuration is a charge $k$
Kaluza-Klein monopole solution, which is known to be T-dual to k
NS5-branes that do not wrap the dualized circle (see, e.g., \cite{Ton}
and references therein).

If we restrict to a linking 2-sphere, we obtain an isomorphism of the 
twisted K-theories of Lens spaces $L(1,p)=S^{3}/\ZZ_p$
\begin{equation}
  K^i(L(1,j),H=k)\cong K^{i+1}(L(1,k),H=j) \,.
\end{equation}
We recall that $L(1,p)=S^{3}/\ZZ_p$ is the total space of 
the circle bundle over the 
2-sphere with Chern class equal to $p$ times the generator 
of $H^2(S^2,\ZZ)\cong \ZZ$. Note that $L(1,1) = S^3$ and $L(1,0)
= S^2 \times S^1$. 

In the case of a single NS5-brane, $j=1$, the total space of the
circle bundle over the linking 2-sphere is a 3-sphere, the group
manifold of $SU(2)$.  Thus we obtain an isomorphism
\begin{equation}
  K^i(SU(2),H=k)\cong K^{i+1}(L(1,k),H=1) \,,
\end{equation}
between the K-theory of $SU(2)$, twisted by $H=k\in 
H^3(S^3,\ZZ)$ and the (parity shifted) K-theory of the Lens 
space $L(1,k)$ twisted by only one unit.  

The special case of string theory on a 7-manifold crossed
with the 3-torus $T^3$ with $k$ units of $H$-flux on the $T^3$ was 
considered in Sect.~\ref{nil}.  This
is a trivial circle bundle over $T^2$, and so $g=1$ and $j=0$.  Using
Eqn.~(\ref{eqAf}), T-duality along any circle yields a circle bundle
over $T^2$ with Chern class $k$ and no $H$-flux.  The total space of
this bundle, in agreement with the literature, is just the $k$th
nilmanifold.

An example along the lines of that in Ref.~\cite{DLP} is $\twob$ on
$AdS^3\times S^3\times T^4$ with $N$ units of $G_3$-flux supported on
the $S^3$.  The 3-sphere is a circle bundle over $S^2$ with Chern
class $j=1$ and one may T-dualize this fiber.  The Chern class is
converted into $H$-flux, and because we began with no $H$-flux the
resulting bundle is trivial.  This leaves type $\twoa$ on $AdS^3\times
S^2\times S^1\times T^4$.  There is now one unit of $H$-flux supported
on the $S^2\times S^1$, as a result of the Chern class of the original
bundle.  The isomorphism of K-groups exchanged $H^2$ and $H^3$ and so
the $G_3$-flux becomes $G_2$-flux.  Thus we find
\begin{equation}
  \int_{S^2\times S^1} H=1\,,\qquad
  \int_{S^2} G_2=N\,.
\end{equation}
The large $N$ duality to a 2$d$ conformal field theory is much more
mysterious in this framework, even the R-symmetry is nontrivially
encoded in the geometry.

% ------------------------------------------------------------------------
\subsection{Bundles over $\rpt$}

In this section we will consider T-dualities of the two circle bundles
over $\rpt$.  To obtain the rest of the nonorientable 3-manifolds
which are circle bundles, one need only connect sum the $\rpt$ with a
Riemann surface, which, as above will add factors of $\Z^{2g}$ which
will play no role.  However the nonorientable cases are more difficult
to adapt to string theory because we cannot make a consistent
background for type $\two$ by simply (topologically) crossing them
with a 7-manifold, as the total space will continue to be
nonorientable.  To make a consistent string theory background from
this example one has several choices.  For example, one may consider
an orientifold projection, or one may consider a topology which is
only locally this example crossed with a 7-manifold.  In the first
case, complex twisted K-theory will no longer be the K-theory which
classifies fluxes and branes.  In the second, the relevant complex
K-theory will not simply be the tensor of the K-theory that we find
below with that of the 7-manifold.  So in either case, adapting the
results below to classify fluxes in a string background is less
trivial than for the other examples of this note. However this example
does illustrate that the twisted K-theory isomorphism appears to work
when $H$ is torsion and also for nonorientable manifolds (although,
strictly speaking, in the discussion upto now we have assumed the
$S^1$-bundle to be orientable).

To classify bundles on $\rpt$, we must first know its 
$\Z$-valued cohomology:
\begin{equation}
  H^0(\rpt,\ZZ)=\ZZ\,,\qquad H^1(\rpt,\ZZ)=0\,,\qquad
  H^2(\rpt,\ZZ)=\Z_2\,.
\end{equation}
T-duality interchanges the Chern class with the $H$-flux, both of
which are zero, and so takes the trivial bundle with no $H$-flux to
itself.  It interchanges $K^0$ and $K^1$, which is consistent with the
fact that they are isomorphic.

We next consider the trivial bundle with 1 unit of $H$-flux.  The cup
product of this $H$-flux with $k\in H^0(\rpt\times S^1)=\Z$ is $k\in
H^3(\rpt\times S^1)=\Z_2$ and so is zero if $k$ is even and one if $k$
is odd.  Thus the subset of $H^0$ that is in the kernel of $H\cup$
consists of the even integers $2\Z\cong \Z$ which are isomorphic to
the integers.  The rest of the cohomology is automatically in the
kernel.  The image consists of $H^3$, and so the quotient of the
kernel by the image is
\begin{eqnarray}
K^0(\rpt\times S^1,H=1)&=&2H^0\oplus H^2=\Z\oplus\Z_2\nonumber\\
K^1(\rpt\times S^1,H=1)&=&H^1\oplus H^3/H^3=\Z. \label{trivb}
\end{eqnarray}
The T-dual is obtained by interchanging the Chern class of the bundle,
which is zero, with $H$, which is one.

The result is the nontrivial bundle with no $H$-flux.  A simple
construction of this nontrivial bundle is as follows.  It is the
nontrivial $S^2$ bundle over $S^1$.  That is to say, begin with the
3$d$ cylinder $S^2\times I$, where $I$ is the interval.  Glue the
$S^2$'s at the two ends of the cylinder together by attaching each
point on the $S^2$ to its antipodal point $(x,0)\sim(-x,1)$, as one
would construct the Klein bottle in the case of a 2$d$ cylinder.  To
see that the resulting space is $E$, an $S^1$ bundle over $\rpt$,
notice that there is an $S^1$ action given by moving along the circle
which we constructed by gluing together the two ends of the interval.
If one begins at $(x,0)$, one arrives later at $(x,1)\sim (-x,0)$ and
later at $(-x,1)\sim (x,0)$ once again.  Thus the space of orbits of
this circle action is just the 2-sphere with $x$ and $-x$ identified.
As desired, this is $\rpt$.  The projection map $E\rightarrow \rpt$
identifies each orbit with the corresponding point in $\rpt$.

We find the homology of $E$ analogously to the case of the 2$d$ Klein
bottle.  The circle generates $H_1(E,\ZZ)=\Z$.  The two-sphere is the
generator $x\in H_2$, but it gets identified with its mirror image,
and so $x\sim -x$ because the antipodal map negates the orientation of
even dimensional spheres.  This yields the relation $2x=0$ and so
$H_2(E,\ZZ)=\ZZ_2$.  The space is not orientable and so the top
homology class vanishes $H_3(E,\ZZ)=0$.  The universal coefficient
theorem allows us to find the cohomology of $E$
\begin{equation}
  H^0(E,\ZZ)=\Z\sp H^1(E,\ZZ)=\Z\sp H^2(E,\ZZ)=0\sp H^3(E,\ZZ)=\Z_2\,.
\end{equation}
The T-dual of the trivial bundle with $H$-flux is $E$ with no flux, 
and so the twisted K-theory is the untwisted K-theory
\begin{equation}
  K^0(E)=H^0(E,\ZZ)\oplus H^2(E,\ZZ)=\Z\,,\quad
  K^1(E)=H^1(E,\ZZ)\oplus H^3(E,\ZZ)=\Z\oplus\Z_2\,.
\end{equation}
As desired, $K^0$ and $K^1$ are the same as the K-groups $K^1$ and $
K^0$ of the T-dual in Eqn.~(\ref{trivb}).

There is one more case.  The nontrivial bundle $E$ may support one 
unit of $H$-flux.  Taking the cohomology with respect to the cup 
product by $H$ proceeds identically to the case of the trivial 
bundle discussed above, and we find
\begin{equation}
  K^0(E,H=1)= 2H^0\oplus H^2=\ZZ\,,\quad
  K^1(E,H=1)= H^1 \oplus H^3/H^3=\ZZ\,. 
\end{equation}
These are the same K-groups as those found in (\ref{trivb}) except 
that $H^2(E,\ZZ)=0\neq H^2(\rpt\times S^1,\ZZ)=\Z_2$ and so $K^0$ does not 
contain a $\Z_2$-factor here.  This is crucial, as it means that 
$K^0(E,H=1)=K^1(E,H=1)$.  This configuration is self-dual under 
T-duality, interchanging $K^0$ and $K^1$.

% -------------------------------------------------------------------------
\section{Application: Circle bundles over $\rp^n$} \label{rpnsec}

In general calculating the twisted K-theory of high-dimensional manifolds is 
quite difficult as many of the differentials of the Atiyah-Hirzebruch 
spectral sequence for twisted K-theory are not known.  Except for the 
$H$-term in $d_3$ used above, these differentials $d_{2k+1}$ take $even$ or 
$odd$ cohomology classes to the torsion part of $odd$ or $even$ 
cohomologies.  As we will see, the odd cohomology classes of $\rp^n$ do 
not contain any torsion, and so no differentials have an image in odd 
cohomology.  Furthermore the only odd cohomology class that is 
nonvanishing is the top-dimensional one, which is automatically 
annihilated by all differentials, and so all odd dimensional cohomology 
is in the kernel of the differentials.  Thus, except for the $H\cup$ term 
used above, all of the differentials act trivially on the cohomology of 
$\rp^n$.  No extra complication is introduced by crossing with a circle, 
and the nontrivial circle bundle is in fact even simpler.  The result is 
that all K-groups in this subsection can be found by taking the elements 
of the cohomology that are annihilated by $H$ and quotienting by those 
that are cup products with $H$, just as in the three-dimensional case.

As explained above, an additional complication arises in the case of 
manifolds of dimension greater than $3$.  The spectral sequence does not 
necessarily yield the desired twisted K-groups, but only an associated graded 
algebra.  To find the K-groups, in general one must then solve an extension
problem.  We will see that in this set of examples T-duality maps bundles
with a nontrivial extension problem to bundles with a trivial extension 
problem, and so T-duality will provide the extension problem's solution.

It will prove to be convenient to treat the case of odd and even $n$ 
separately.  For example, the $\rp^{2m+1}$'s are orientable and the 
$\rp^{2m}$'s are not.  It is therefore the odd $n$ cases that are 
directly applicable to consistent type $\two$ 
string theory compactifications.  
The nontrivial integral cohomology groups are 
\begin{equation}
  H^0(\rp^{n},\ZZ)=\Z\sp H^{2p}(\rp^{n},\ZZ)=\Z_2\,,\quad 
  H^{2m+1}(\rp^{2m+1},\ZZ)=\Z\,,\quad
  p=1,\ldots,\lfloor \frac{n}{2}\rfloor\,.
\end{equation}
The cohomology of the trivial circle bundle is similarly
\begin{align}
  & H^0(\rp^{n}\times S^1,\ZZ)=H^1(\rp^{n}\times S^1,\ZZ)=\Z\sp 
  H^{q}(\rp^{n}\times S^1,\ZZ)=\Z_2\sp q=2,\ldots,n-1 \,, 
     \nonumber\\
  & H^{2m}(\rp^{2m}\times S^1,\ZZ)=H^{2m+1}(\rp^{2m}\times S^1,\ZZ)
  =\Z_2 \,, \nonumber\\
  & H^{2m+1}(\rp^{2m+1}\times S^1,\ZZ)=\Z\oplus\Z_2\sp
  H^{2m+2}(\rp^{2m+1}\times S^1,\ZZ)=\Z \,.
\end{align}
where we have assumed that $n>1$, thus losing the case of $\rp^1$ in which 
no nontrivial fibrations are possible.

Possible twists are elements of the third cohomology group
\begin{equation*}
H^3(\rp^n\times S^1,\ZZ)=\left\{
  \begin{matrix} \Z\oplus\Z_2\hsp{.2}&\hsp{.2}\textup{if $n=3$}\,,\cr 
  \hsp{-.5}\Z_2&\hsp{.2}\textup{if $n\neq 3$} \,.
  \end{matrix}
  \right.\,. 
\end{equation*}  
The extra $\Z$ in the special case of $\rp^3$ consists of classes in
$H^3(\rp^3)=\Z$, and not in $H^2(\rp^3,\ZZ)\otimes H^1(S^1,\ZZ)=\Z_2$.
Therefore when integrated over the circle $H$-twists in this $\ZZ$ are
trivial, and do not change the topology of the T-dual manifold.  Of
course, it is possible that $H$ is the sum of such a class with the
nontrivial element of the $\ZZ_2$, that is $H=(k,1)$.  In this case it
will be a critical consistency check of our conjecture that the T-dual
manifold also have a subgroup $\ZZ\subset H^3(E,\ZZ)$ so that there
may be a T-dual flux $\hat{H}=(k,0)$.  We will see that the cohomology
of the T-dual does in fact have such a subgroup.

We begin again with the case of vanishing $H$-flux.  In this case the
K-theory is simply the cohomology
\begin{align} \label{rpntriv}
  & K^0(\rp^{2m}\times S^1)=\bigoplus_p H^{2p} =
           \Z\oplus\Z_2^{m}\,,\quad
  K^1(\rp^{2m}\times S^1)=\bigoplus_p H^{2p+1} = \Z\oplus\Z_2^{m} 
      \,,\nonumber\\
  & K^0(\rp^{2m+1}\times S^1)=\bigoplus_p H^{2p} = 
     \Z^2\oplus\Z_2^{m}\,,\quad
  K^1(\rp^{2m+1}\times S^1)=\bigoplus_p H^{2p+1} = \Z^2\oplus\Z_2^{m}\,. 
\end{align}
As the Thom isomorphism or equivalently here the K\"unneth theorem
guarantees, in each case $K^0\cong K^1$ and so T-duality on the circle 
simply acts by interchanging classes in these two K-groups.  As a check on 
these results, one may recall that $\rp^{2m+1}$ is a circle bundle over
$\CP^m$ with two units of Chern class and one may T-dualize about that 
circle.  This yields $\CP^m\times S^1$ with $H=2$.  
The twisted K-theory of $\RP^{2m+1}\times S^1$ is then just the cohomology of 
$\CP^m\times T^2$, which consists of $\Z^2$ for 
each group, quotiented by $H$.  A quick calculation shows 
that these twisted K-groups agree with their T-duals in 
Eqn.~\eqref{rpntriv}.

If we turn on nontrivial $H$-flux in the $\Z_2\subset H^3$ then the
twisted K-theory will be the kernel of $H\cup$ quotiented by its
image.  This flux cups nontrivially on even cohomology groups, taking
each to the $Z_2$ torsion part of the odd group three dimensions
higher.  In particular all torsion odd cohomology groups are in the
image and so are quotiented out of the K-theory.  Only even elements
of the even-dimensional cohomology groups are in the kernel, which
means only the zero elements of the torsion groups, and $2\Z\cong\Z$
in $H^0$.  All of $H^{2m-1}$ and $H^{2m}$ is in the kernel for
dimensional reasons.  In sum, the twisted K-theory is
\begin{align} \label{guess}
  & K^0(\rp^{2m}\times S^1,H=1)\stackrel{?}{=}2H^{0}\oplus H^{2m} = 
     \Z\oplus\Z_2\nonumber\,,\\
  & K^1(\rp^{2m}\times S^1,H=1)\stackrel{?}{=}2H^{1} = \Z\nonumber\,,\\
  & K^0(\rp^{2m+1}\times S^1,H=1)\stackrel{?}{=}2H^{0}
    \oplus H^{2m}\oplus H^{2m+2} =\Z^2\oplus\Z_2\nonumber\,,\\ 
  & K^1(\rp^{2m+1}\times S^1,H=1)\stackrel{?}{=}2H^{1}\oplus H^{2m+1} = 
  \Z^2\,.
\end{align}
The question marks indicate that there is a nontrivial extension problem to 
solve here, which will be solved later by imposing our T-duality conjecture
and also argued from the explicit construction of our isomorphism.
As noted above, in the case of $\rp^3\times S^1$, one may also add $m$
units of nontorsion $H$-flux.  In this case the $\Z^2$'s above are
replaced by $\Z_m$'s.

The T-dual is the nontrivial circle bundle $E_n$ over $\rp^n$, which
as above is an $S^n$-bundle over the circle made from $S^n\times I$
via the gluing $(x,0)\sim(-x,1)$.  Notice however that in the case of
odd $n=2m+1$ the map $x\mapsto -x$ is homotopic to the identity, and
so for odd $n$ the T-dual space is $S^1\times S^{2m+1}$.  The
cohomology is found as in the $\rpt$ case for $n$ even and by
K\"unneth for $n$ odd to be
\begin{align}
  & H^0(E_n,\ZZ)=H^1(E_n,\ZZ)=\Z\,,\qquad
    H^{2m}(E_{2m+1},\ZZ)=\Z\,,\nonumber \\
  & H^{2m+1}(E_{2m},\ZZ)=\Z_2 \,,\qquad
    H^{2m+1}(E_{2m+1},\ZZ)=\Z.
\end{align}
This allows for $H$-flux only in the cases of $\rpt$, treated above,
and also $\rp^3$.  $H^3(\rp^3,\ZZ)=\Z$ and so the $H$-flux may assume any
integer value, which is reassuring as the T-dual also allowed for an
extra integer in the definition of the $H$-flux.  These two integers
must agree.

Thus we need consider only the case of vanishing $H$-flux, and so the
K-groups are just the cohomology groups
\begin{align}
  & K^0(E_{2m})=H^0=\Z\,,\qquad
  K^1(E_{2m})=H^1\oplus H^{2m+1}=\Z\oplus\Z_2 \,,
     \nonumber\\
  & K^0(E_{2m+1})=H^0\oplus H^{2m}=\Z^2\,,\qquad
  K^1(E_{2m+1})=H^1\oplus H^{2m+1}=\Z^2\,.
\end{align}
These groups are all consistent with their T-duals as calculated in 
Eqn.~\eqref{guess}, except for
\begin{equation}
  K^1(E_{2m+1})=\Z^2\quad \neq\quad \Z^2\oplus\Z_2=
  K^0(\rp^n\times S^1,H=1)\,.
\end{equation}
{}From this we infer that the associated graded algebra and the K-group
are in fact different in this case.  The relevant extension
problem is
\begin{equation}
  0\longrightarrow \Z_2\longrightarrow K^1(E_{2m+1}) 
  \longrightarrow \Z^2\longrightarrow 0 \label{eprob}
\end{equation}
which admits $\Z^2$ as a solution as well as $\Z^2\oplus \Z_2$, the
solution which we assumed above.  Our T-duality conjecture appears to
predict that the desired solution is $\Z^2$.

This solution to the extension problem can be inferred topologically
from our construction of the isomorphism of twisted K-groups.  The
fibered product of our two circle bundles is $S^{2m+1}\times S^1\times
\hat S^1 $ and it fits into the following commutative diagram
\begin{equation} 
\xymatrix @=8pc @ur { \rp^{2m+1}\times S^1 \ar[d]_{\pi} & 
S^{2m+1}\times S^1\times \hat S^1  \ar[d]_{\hat p} \ar[l]^{p} \\ \rp^{2m+1}
& S^{2m+1}\times \hat S^1  \ar[l]^{\hat \pi}}
\end{equation}
Recall that the top cohomology group of our trivial $\rp^{2m+1}$
bundle is $H^{2m+2}(\rp^{2m+1}\times S^1,\ZZ)=\Z$.  
This is the Poincar\'e dual
of a point $x$.  The key realization is that the preimage of this
point $p^{-1}(x)$ is a circle which wraps $\hat S^1 $ \textit{twice}.
This is because the projection map $p$ projects to the orbits of a
circle which simultaneously wraps $\hat S^1 $ and acts on the
$S^{2m+1}$ via a nonvanishing vectorfield scaled such that after
wrapping $\hat S^1 $ once, one arrives at the antipodal point in
$S^{2m+1}$.  Thus the orbit only closes after wrapping $\hat S^1 $ a
second time.

Our isomorphism, acting now on integral homology, takes $x$ to
$\hat{p}p^{-1}(x)$, which again wraps $\hat{S}^1$ twice.  The
tensoring with the Poincar\'e bundle is trivial because $p^{-1}(x)$ does
not wrap $S^1$.  In sum, we have found that
\begin{equation}
  T_*:H_0(\rp^{2m+1}\times S^1,\ZZ)=\Z\quad \longrightarrow \quad
  H_1(S^{2m+1}\times \hat S^1,\ZZ )=\Z\ :\ 1\mapsto 2\,.
\end{equation}
This means that the class $1\in H_0(\rp^{2m+1}\times S^1,\ZZ)$ actually
corresponds to the class $2$ in twisted K-theory, which is only
consistent if the solution to the extension problem \eqref{eprob} is
given by
\begin{equation}
  K^1(E_{2m+1})=\Z^2 \,.
\end{equation}

% =========================================================================
\section{Anomalies} \label{AnSec}
\subsection{Quotients of $AdS^5\times S^5$}

A more nontrivial check of our conjecture (\ref{eqAf}) comes in its
application to circle bundles on $\CP^2$.  
We have $H^2(\CP^2)=\Z$, and so again
circle bundles are parametrized by a single integer $j$.  The total
space of such a bundle is the Lens space $L(2,j)$, i.e.\ the
nonsingular quotient $E=S^5/\Z_j$, when
$j\neq 0$, and $E=\CP^2\times S^1$ when $j=0$.  The nonvanishing
integral cohomology groups are (see, e.g., \cite{BT})
\begin{align}
  & H^{0\leq p\leq 5}(\CP^2\times S^1)=\Z \,, \\
  & H^0(L(2,j),\ZZ)=H^5(L(2,j),\ZZ)=\Z\,,\qquad  
  H^2(L(2,j),\ZZ)=H^4(L(2,j),\ZZ)= \Z_j \,. \nonumber
\end{align}
Thus $H$-flux is only possible for the trivial bundle $j=0$, as the
nontrivial bundles have trivial third cohomology.  In the case of the
trivial bundle, the cup product with the $H$-flux maps $H^0$ to $H^3$
and $H^2$ to $H^4$ while $H^1, H^3$\ and $H^5$\ are all in
$ker(d_3=H\cup)$.  The next differential in the spectral sequence,
$d_5$, may act nontrivially on the cohomology ring,\footnote{Whether it does
depends on an ill-defined division by 2 in Ref.~\cite{uday}.} but is
trivial on the kernel of $d_3$ and so does not affect the twisted
K-theory of $\CP^2\times S^1$.

T-duality relates the trivial bundle with $H=j$ to the bundle with
first Chern class $j$ and no flux.  The twisted K-theory of the former
is
\begin{align}\label{casea}
  K^0(\CP^2\times S^1,{H=j})&=H^4(\CP^2\times S^1,\ZZ)=\Z \,, \nonumber\\
  K^1(\CP^2\times S^1,{H=j})&=H^1\oplus H^3\oplus H^5 /(jH^3\oplus jH^5)=
  \Z\oplus\Z_j{}^2\,. 
\end{align}
In the latter case $H$ vanishes and so the K-groups are just the 
cohomology groups
\begin{align}
  & K^0(L(2,j))=H^0\oplus H^2\oplus H^4=\Z\oplus\Z^2_j\,, \nonumber\\
  & K^1(L(2,j))=H^1\oplus H^3\oplus H^5=\Z\,. \label{caseb}
\end{align}
And so we see that cases (\ref{casea}) and (\ref{caseb}) differ 
by the exchange of $K^0$ and $K^1$ as desired.

Of course such T-dualities are interesting because $\twob$ string theory
on $AdS^5\times S^5$ is comparably well understood.  This $j=1$
example of the above T-duality was first studied in Ref.~\cite{DLP}
where it was observed that the spacetime on the $\twoa$ side is not
$spin$, making the duality quite nontrivial.

The resulting RR fluxes are easily computed.  If we start with $N$
units of $G_5$-flux supported on $L(2,j)$ in $\twob$, then in $\twoa$ there
will be $N$ units of $G_4$-flux supported on $\CP^2$ and $j$ units of
$H$-flux supported on $H^2(\CP^2,\ZZ)\otimes H^1(S^1,\ZZ)$.

% -------------------------------------------------------------------------
\subsection{Gravitino Anomalies Before and After}

One might worry that type $\twoa$ string theory (and also its M-theory
lift) on a non-$spin$ manifold is inconsistent, because the gravitino
requires a spin structure to exist.  There is no such anomaly on the
$\twob$ side, whose space-time is the $spin$ manifold $AdS^5\times S^5$,
thus it is a critical check of this duality that the anomaly be
cancelled on the $\twoa$ side.

The authors of Ref.~\cite{DLP} have shown that the anomaly is in 
fact cancelled.  This cancellation is a result of the 11$d$ 
supergravity coupling
\begin{equation}
  \L_{11d}\supset \overline{\Psi} G_4 \Psi 
\end{equation}  
of the gravitino $\Psi$ to the 4-form fieldstrength $G_4$.  
Dimensionally reducing away the M-theory circle and $S^1_{\twoa}$ one 
finds, among other terms, the 9-dimensional coupling
\begin{equation}
  \L_{9d}\supset \overline{\Psi} F_2 \Psi 
\end{equation}  
identifying the gravitino as a fermion charged under a $U(1)$ gauge symmetry.  
The anomaly should be independent of the high energy physics 
such as the massive
KK-modes which are uncharged under this $U(1)$.

Such a fermion may be consistent on a manifold $M$ that is not $spin$, 
but is merely $spin^\CC$ if the second Stiefel-Whitney class $w_2(M)$ 
is equal to twice the fermions charge $Q$ multiplied by the Chern 
class of the bundle
\begin{equation}
  w_2(TM)=2Q c_1(E)\,. \label{cond}
\end{equation}
The right hand side of this equation is naturally an 
element of $H^2(M)$ with integral coefficients. The left hand side of 
course is an element of cohomology with $U(1)$ coefficients, but due to the $spin^\CC$ 
condition it also lifts to integral cohomology.   To find $c_1(E)$, recall that, according to the $E_8$ interpretation,
the fibers of the $U(1)$ bundle are just the circle $S^1_{\twob}$ that
appears in type $\twob$.  Thus the Chern class is $j=1$, more precisely,
it is the generator of $H^2(\CP^2,\ZZ)$.  The second Stiefel-Whitney class
of the $\twoa$ spacetime is the same class, and so the anomaly
cancellation condition (\ref{cond}) is only satisfied if $Q$ is
half-integral.

In Ref.~\cite{DLP} it was concluded that $Q$ is in fact half-integral
and so the anomaly vanishes on the $\twoa$ side.  To see this, perform a
gauge transformation by an angle of $2\pi$.  This contributes a phase
to the gravitino's wavefunction
\begin{equation}
  \Psi\longrightarrow e^{2\pi Q}\Psi\,.
\end{equation}
To calculate this phase, we look at the $\twob$ side.  This is a rotation
of the $\twob$ circle over $2\pi$, and so corresponds to
transporting the gravitino around the circle.  If we chose the
supersymmetric spin structure on the circle then the gravitino's phase
acquires a $-1$, and so $Q$ is half-integral as required.  It is
interesting that the matching of anomalies required us to choose the
supersymmetric spin structure on the circle about which we T-dualized,
if we had not then the result may not have been $\twoa$ but possibly 
type-$0$ \cite{DGHM}.

% -------------------------------------------------------------------------
\subsection{The Gravitino Anomaly in the General Case}

We have found that the $H$-field arising from our T-duality cancels
the gravitino anomaly on the $\twoa$ side, so that the $\twoa$ theory is
consistent.  It is a critical test of our proposal \eqref{eqAf} that
the two sides of the duality be consistent and inconsistent at the
same time.  That is, the gravitino anomalies must match on the two
sides in general.  To see that they do, we extend the above argument
to the general case.  We will begin with the case in which there is no
$H$-flux on the $\twob$ side, and so a trivial bundle in $\twoa$.

As there is no $H$-flux on the $\twob$ side, the gravitino anomaly is
entirely determined by the second Stiefel-Whitney class of the
$S^1_{\twob}$ bundle $E$
\begin{equation} \label{ab}
\textup{Anomaly}_{\twob}=w_2(TE)=w_2(M^9)+w_2(E)=w_2(M^9)+c_1(E) \,, 
  {\textup\ {\rm mod}\ 2}
\end{equation}
where $w_2(TE)\subset H^2(E)$ is the Stiefel-Whitney class of the
tangent space to $E$, whereas $w_2(E)\subset H^2(M^9)$ and
$c_1(E)\subset H^2(M^9)$ are characteristic classes of the $S^1$
bundle over the base $M^9$.  In the case of $L(2,j)=S^5/\Z_j$ 
this anomaly is
$1+j$ and so the $\twob$ side is anomalous when $j$ is even.

To compute the anomaly on the $\twoa$ side we will first dimensionally
reduce away the trivially fibered circle $S^1_{\twoa}$.  
If our T-duality conjecture is
correct this will be $\twob$ reduced to 9-dimensions and so the anomalies
will match.  To check that it does, notice that the anomaly for a
$U(1)$ charged fermion in 9 dimensions is given by (\ref{cond})
\begin{equation} \label{aa}
  \textup{Anomaly}_{\twoa}=w_2(M^9)+ 2Q c_1(E) {\textup\ {\rm mod}\ 2}
\end{equation}
where $E$ is now interpreted as our gauge bundle, although the $E_8$
description tells us that it is the same $E$ as we encountered on the
$\twob$ side.  If we again take the 
supersymmetric spin structure then by
the same argument we conclude that $Q$ is half-integral and so the
anomalies (\ref{aa}) and (\ref{ab}) as computed in type $\twoa$ and $\twob$
agree.

To extend this argument further, to the general case in which there is
$H$-flux before and the T-duality, one need only observe that the
total anomaly in both cases is the second Stiefel-Whitney class of the
sum of the two circle bundles.  Thus they are both the sum of
$w_2(M^9)$ plus $w_2$ of the two circle bundles, where the fact that
$Q$ is half-integral for the chosen spin structure has been used to
rewrite one Chern class as a Stiefel-Whitney class.  As both anomalies
are given by the same formula, they agree.  It is suggestive
(mysterious) to rewrite the anomaly as $w_2$ of the fibered product.
One may then include the $\twoa$ coupling of the gravitino to $G_2$ to
conclude that the total anomaly is $w_2$ of the F-theory 12-manifold.

% --------------------------------------------------------------------------
\subsection{The $G_4$ Quantization Condition in M-theory}

In Ref.~\cite{Wit} Witten showed that when the spacetime $Y^{11}$ is $spin$ the M-theory four-form
obeys the twisted flux quantization condition which we heuristically write as
\begin{equation} \label{witten}
  G_4= \textstyle{\frac14} p_1(TY^{11})\textup{\ mod\ 2}.
\end{equation}
For an interpretation of these divisions, we refer the reader
to the original paper.  While the first Pontrjagin class
$p_1$ of the tangent space may always be canonically divided by two,
the division by 4 in \eqref{witten} is canonical because $Y^{11}$ is
$spin$.

As explained in \cite{DLP}, $G_2$ vanishes in the above example of $\twoa$
on $AdS^5\times \CP^2\times S^1$ and so the M-theory topology is
$AdS^5\times \CP^2\times T^2$, which is not $spin$.  Therefore the
above divison by two may not exist.  In fact, the $G_4$ flux in this
example is a cup product of the generator of $H^2(T^2,\ZZ)=\Z$ and so does
not satisfy the twisted quantization condition \eqref{witten}.

Instead we see that when M-theory is compactified on a 2-torus 
$T^2$ the shifted quantization condition is
\begin{equation}
  \int_{T^2} G_4 = w_2(TM^9)\textup{\ mod 2}\,.
\end{equation}
This equation may well generalize to 2-torus bundles, and possibly the
2-torus may be replaced by any 2-manifold.  When the spacetime is not
of such a form, perhaps the anomaly-cancellation used above cannot
work and so the 11-dimensional spacetime must be $spin$, and thus
Eqn.~(\ref{witten}) is applicable.  Nonetheless, it may be interesting
to find a single formula that works in all of the cases.

% ======================================================================
\section{Concluding remarks} \label{GenSec}

We have conjectured that any orientable circle bundle is T-dual to
another circle bundle, where the Chern class of each bundle is the
integral of the T-dual $H$-flux over the dual circle.  As evidence, we
have provided physical motivation in a number of special cases and
have seen that this definition of T-duality always leads to the
desired isomorphism of the desired twisted K-theories with a shift in
dimension by one.

However to be certain that this isomorphism of twisted K-theory is a
duality of the full string theory in the most general cases one
requires more powerful methods.  The most obvious choice is the
$\sigma$-model on $E\times_M\hat E$ program of \cite{RV} and later
\cite{AABL}.  This approach has been used to find that nontrivial
bundles are dual to a singular $B$-flux.  Thus it may be possible to
compute the corresponding $H$-flux and verify that it obeys our
conjecture.  This calculation would then need to be extended to the
case in which $H$ is nontrivial both before and after the T-duality.

This approach may allow a number of other open problems to be tackled
directly.  An obvious one is the generalization of the results 
of this paper to
higher-dimensional torii.  The obstruction that we conjecture exists
when the integral of $H$ over a subtorus is nontrivial may be visible
directly in such an approach.  In the present approach, the
obstruction is mysterious because the S-dual to an obstructed
T-duality is the T-duality of a 2-torus supporting $G_3$-flux.  Such a
T-duality is perfectly legitimate, and leads to $G_1$-flux.  Thus one
may suspect that the forbidden T-duality of a 2-torus with $H$-flux
yields the S-dual of a configuration with $G_1$-flux.  The S-duals of
such configurations have been described extensively in the literature,
but unfortunately the descriptions tend not to agree.  One common
feature among papers that claim that such a duality makes sense is
that the dilaton ceases to be globally defined, which may explain why
we have difficulty understanding such a theory.

Another obvious generalization is that we may allow our circle fibers
to degenerate.  This would then include examples such as mirror
symmetry.  While the $\sigma$-model approach may be promising here,
the traditional approach to this subject \cite{Phases,HV} suggests
that a linear sigma model which flows in the IR to the conformal
theory may provide a much more practical tool for this and the
previous generalizations.

The generalization to the equivariant case appears to be
straightforward, and we hope to come back to this 
in future work.  More intruiging is the
extension from $U(1)$-bundles to non-abelian bundles yielding
nonabelian dualities.  Such bundles are also treated in \cite{RR}
although the results are much more limited.

There are a number of more tangentially related applications and open
problems.  As mentioned above, the shifted quantization condition of
$G_4$ in the non-$spin$ case is still unknown in general, although in
the Riemann-surface bundle case the contribution above may be the
entire condition.  The T-duality above between $\rp^3\times S^1$ with
$H$-flux and $S^3\times S^1$ may be dimensionally reduced to a
7-dimensional duality of gauged supergravities.  This may relate an
$SU(2)$ symmetry to an $SO(3)$ symmetry with a $\Z_2$ Wilson line
activated.

Perhaps the most mysterious aspect of this realization of T-duality is
the connection to F-theory.  Consistency seemed to require that the
12-manifold be $spin$, as if it were inhabited by fermions despite the
lack of a single-time 12$d$ SUSY algebra.  More significantly, the
$\sigma$-model approach introduces an auxilliary dimension as an
intermediate step, and that step seems to be a kind of $\sigma$-model
on the fibered product, which is F-theory compactified on the M-theory
circle.  Could this mean that F-theory is a theory after all?

\bigskip\bigskip

\noindent 
{\bf Acknowledgements}

\noindent  
We would like to thank Nick Halmagyi, Michael Schulz and Eric Sharpe
for help and crucial references.  JE would like to thank the
University of Adelaide for hospitality during this project.
PB and VM are financially supported by the Australian Research Council
and JE by the INFN Sezione di Pisa.

\noindent

% =========================================================================

% =========================================================================
\end{document}